\documentclass{article}
\usepackage{commands}
\usepackage{xcolor}
\usepackage{authblk}
\usepackage{soul}
\usepackage{lineno}

\newcommand{\bx}{\bm{x}}

\usepackage{xr}
\makeatletter
\newcommand*{\addFileDependency}[1]{
  \typeout{(#1)}
  \@addtofilelist{#1}
  \IfFileExists{#1}{}{\typeout{No file #1.}}
}
\makeatother

\newcommand*{\myexternaldocument}[1]{%
    \externaldocument{#1}%
    \addFileDependency{#1.tex}%
    \addFileDependency{#1.aux}%
}

\myexternaldocument{SI}

\newcommand*\siref[1]{%
    \textcolor{blue}{Supplementary} \cref{#1}}

\title{}

\author[1,2,*]{Àlex Giménez-Romero}
\author[3]{Bernard A. Afful}
\author[4]{Priscilla E. Greenwood}
\author[2]{Manuel A. Matías}
\author[3]{Luis F. Gordillo}

\affil[1]{Department of Ecology and Complexity, Centre d'Estudis Avançats de Blanes (CEAB-CSIC), Blanes, Girona, Spain}
\affil[2]{Instituto de Física Interdisciplinar y Sistemas Complejos (IFISC, CSIC-UIB), Campus UIB, 07122 Palma de Mallorca, Spain}
\affil[3]{Department of Mathematics and Statistics, Utah State University, Logan, Utah}
\affil[4]{Department of Mathematics, University of British Columbia, Vancouver, BC, Canada}
\affil[*]{Corresponding author}

\DeclareUnicodeCharacter{0301}{!!}

\begin{document}

\maketitle

\begin{abstract}
    Vegetation in semi-arid ecosystems frequently organizes into spatially heterogeneous mosaics that regulate ecosystem functioning, productivity, and resilience. These patterns arise from local biological interactions, including facilitation among neighboring plants and competition for limiting resources. Classical theoretical approaches have attributed such organization to scale-dependent feedbacks, predicting regular spatial patterns and abrupt transitions to collapse.
    However, growing empirical and theoretical evidence reveal that environmental variability and demographic stochasticity can fundamentally reshape spatial organization, driving irregular clusters, dynamic mosaics, and gradual rather than catastrophic vegetation declines. In drylands, rainfall variability is a dominant source of environmental forcing: precipitation typically occurs in short, irregular pulses that transiently enhance survival and recruitment before competitive interactions again dominate. Near persistence thresholds, ecosystem dynamics are therefore governed not only by average climatic conditions but also by the timing and spatial coincidence of favorable events. Under these conditions, positive density dependence and local facilitation can critically determine whether vegetation patches persist, expand, or collapse. Here, we develop an individual-based model that integrates intermittent precipitation with local Allee effects to examine how stochastic rainfall shapes spatial organization and persistence. We show that the interaction between pulsed resource availability and density-dependent survival generates irregular cluster structures and strongly modulates extinction risk, with resilience emerging from local spatial covariance and neighborhood density rather than from total biomass alone. These results highlight the importance of individual-level, stochastic processes in determining ecosystem resilience and provide a mechanistic basis for understanding vegetation persistence under increasing climatic variability.
\end{abstract}

\section{Introduction}

Drylands, encompassing arid, semi-arid, and dry sub-humid regions, cover more than 40\% of the terrestrial surface and support a substantial portion of the global human population \cite{assessment2005ecosystems}. These essential ecosystems operate under chronic water scarcity, where vegetation dynamics is driven by infrequent, highly variable rainfall pulses and intense solar radiation \cite{Whitford2019}. Under these harsh conditions, plant communities frequently self-organize into striking spatial patterns—such as bands, spots, labyrinths, and irregular clusters—that are observed across continents, from the ``tiger bush'' stripes of the Sahel to the gap patterns of the Australian Outback and the spot patterns of Namibia \cite{Klausmeier1999, Deblauwe2008}. These patterns play a critical functional role: by concentrating scarce water and enhancing infiltration, they buffer vegetation from aridity and improve survival rates \cite{Rietkerk04,Rietkerk2008,Zhenpeng2023}. Consequently, the formation and persistence of these self-organized structures are closely tied to the resilience of dryland ecosystems \cite{Scheffer2012,Kefi2014,Rietkerk21,Kefi2024}. Therefore, understanding the physiological, ecological, and physical principles that underpin these patterns is a central challenge in ecological theory, particularly for predicting the stability and future trajectory of these critical ecosystems in the face of increasing aridity due to global climate change \cite{Huang2016}.

The prevailing theoretical framework for dryland vegetation patterning is based on Turing-like instabilities, in which spatial structures spontaneously emerge from the symmetry-breaking destabilization of an initially uniform state. Originally, Turing demonstrated this mechanism in a two-species (activator and inhibitor) reaction-diffusion system modeled using partial differential equations (PDEs) \cite{Turing1952}. In this case, stationary spatial patterns arise when the inhibitor diffuses significantly faster than the activator. In vegetation models, these mechanisms are typically represented using reaction–diffusion or advection–diffusion equations for biomass and water, capturing the interplay between local growth and spatial redistribution of limiting resources \cite{Klausmeier1999,Rietkerk2002,Gilad2004,Gilad2007}. Pattern formation arises when water redistribution occurs over spatial scales larger than those of biomass dispersal, or when advective transport along slopes introduces directional asymmetries \cite{Klausmeier1999}. Complementary integro-differential formulations incorporate nonlocal interactions through convolution kernels, allowing more explicit representations of long-range resource redistribution across heterogeneous landscapes \cite{Lefever1997,Martínez-García2013}.

However, the precise mechanisms underlying the emergence of these patterns remain unclear. Traditionally, they have been interpreted through the lens of scale-dependent feedback, whereby the interplay between short-range facilitation and long-range competition drives self-organization \cite{Rietkerk04, Rietkerk2008}. Individual plants locally improve microenvironmental conditions—by enhancing infiltration, shading the soil, or reducing evaporation—thereby promoting establishment and growth in their immediate vicinity, while simultaneously competing for water and nutrients across larger spatial scales through root systems that extend beyond the canopy. Over the last two decades, various formulations of this mechanism have successfully reproduced the characteristic morphologies observed across drylands, including the canonical transition from gaps to labyrinths and spots along aridity gradients \cite{Borgogno2009}, as well as the conditions leading to catastrophic shifts toward desertification \cite{Kéfi2007}. However, recent work has shown that the mechanistic basis of patterning is not unique: spatially extended competition alone, provided it operates over multiple characteristic distances, can yield similar regular and irregular patterns \cite{Martínez-García2013,Martinez-Garcia23}, demonstrating that facilitation is not a strict prerequisite for self-organization.

\begin{figure}[H]
    \centering
    \includegraphics[width=\linewidth]{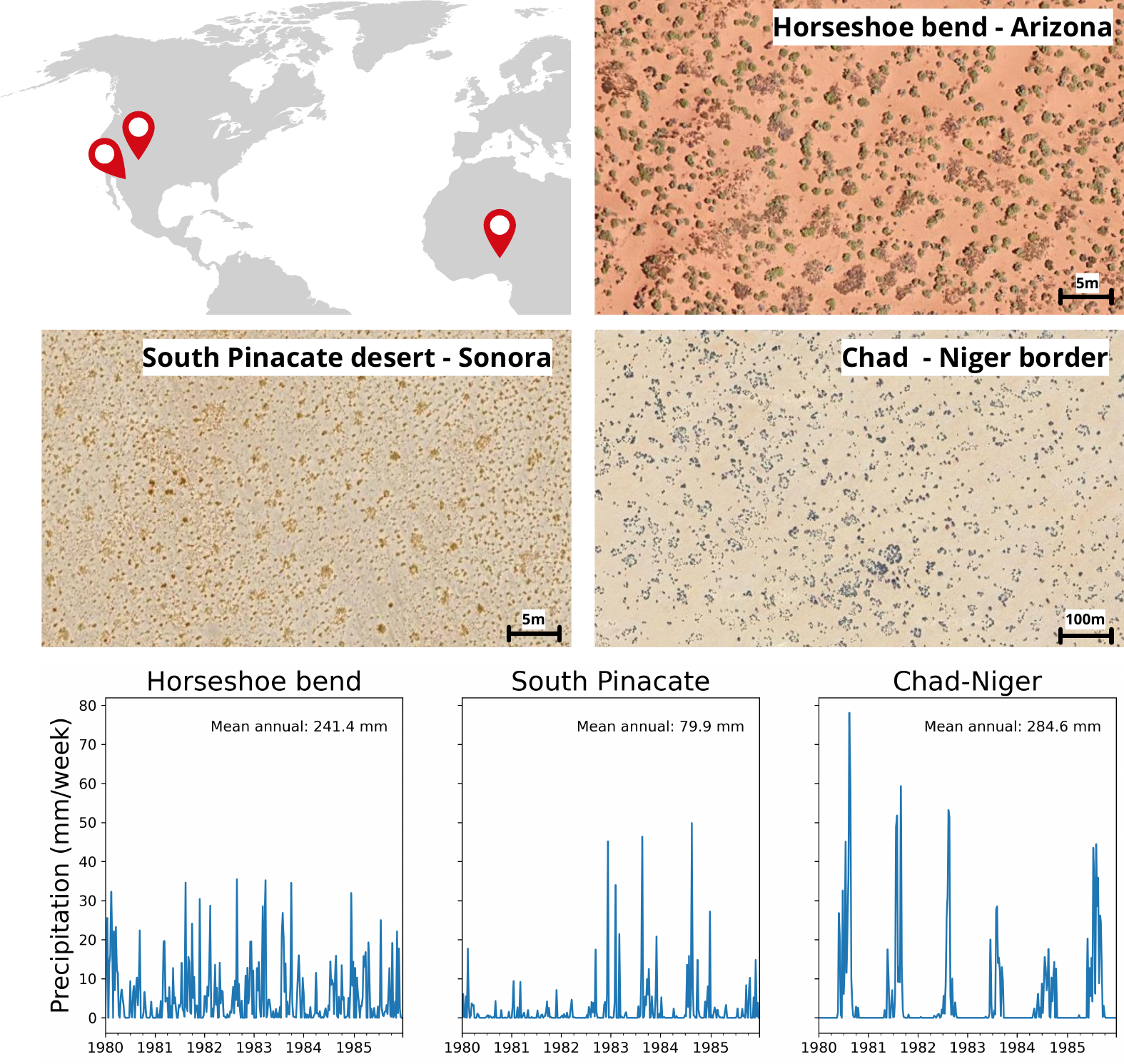}
    \caption{\textbf{Examples of irregular clusters of vegetation}. Horseshoe Bend, Arizona (36.875470º N, 111.504602º W). Pinacate Desert, Sonora (31.534861º N, 114.189449º W). Southern Chad-Niger border (14.695937º N, 13.751921º E).}
    \label{fig:examples_clusters}
\end{figure}

Despite their conceptual contributions, deterministic continuous models face critical limitations when applied to real drylands, as they are mean-field approximations that capture only the leading mechanisms. Lacking spatial heterogeneity and stochasticity in their formulation, these models predict abrupt and catastrophic shifts to bare soil via saddle-node bifurcations. In contrast, the inclusion of demographic noise often transforms these abrupt transitions into progressive declines, weakening pattern robustness and obscuring early warning signals in finite populations \cite{Martínez-García2013,Martinez-Garcia23}. These frameworks also generate highly regular, almost crystalline spatial arrangements that contrast sharply with the distorted and irregular vegetation patterns observed in nature \cite{Kästner2024,Pinto-Ramos2023,Yizhaq2014}. Numerous studies indicate that such discrepancies originate from processes absent in idealized models, including stochastic fluctuations \cite{Odorico2006,Collins2014}, spatial variability in soil–water redistribution \cite{Yizhaq2014}, and heterogeneity in plant mortality or recruitment \cite{Pinto-Ramos2022}. Incorporating these forms of variability not only produces more realistic spatial structures but also may profoundly alter system-level dynamics, frequently replacing deterministic collapses with smooth transitions in average biomass \cite{Pinto-Ramos2022,VillaMartin2015,Yizhaq2016,Yizhaq2014}. Furthermore, classical frameworks typically impose stationary, periodic patterns determined by mean climatic drivers, such as annual rainfall, neglecting the strongly intermittent character of precipitation in semi-arid regions, where inter-event timing varies widely, even among sites with comparable annual totals \cite{Szeles2025} (see \cref{fig:examples_clusters}). Indeed, empirical observations show that vegetation cover fluctuates markedly in synchrony with episodic or seasonal rainfall pulses \cite{Schmidt2000,DeCoca2004,LaPierre2016,Sanz2022}, revealing dynamic vegetation mosaics rather than static structures. Taken together, these advances underscore a central emerging perspective: environmental stochasticity and landscape heterogeneity are not peripheral complications but essential drivers of dryland spatial organization and resilience \cite{Yizhaq2017,Gherardi2019}.

Recent modeling efforts have begun to explicitly incorporate environmental variability to better understand vegetation patterns in drylands. Temporal fluctuations in water availability have been shown to fundamentally reshape spatial organization: intermittent rainfall can induce broad, scale-free patch size distributions through a fluctuating percolation mechanism \cite{martin2020intermittent}, and pulse-driven hydrological dynamics alter the conditions under which patterned states emerge, suppressing diffusion-driven instabilities and sustaining uniform vegetation under precipitation regimes that would otherwise support patterns \cite{Eigentler2020}. Spatial heterogeneity exerts similarly profound effects. Analyses of vegetation systems with spatially varying mortality or environmental stress reveal hysteresis between ordered, high-biomass lattices and disordered, low-biomass cluster mosaics, in agreement with field and remote sensing observations of irregular, non-periodic patterning under increasing aridity \cite{Pinto-Ramos2025}. Complementing these findings, spatial stochastic models that combine Allee effects with competitive interactions have shown that precipitation-frequency–driven transitions generate aggregation patterns, with survival–extinction thresholds tightly linked to spatial randomness and initial configuration \cite{Gordillo23}. Collectively, these studies highlight that stochasticity and heterogeneity—whether in time or space—do more than modulate classical pattern-forming mechanisms; they restructure the bifurcation landscape and generate dynamic, transient, or disordered configurations that fall outside the predictions of deterministic and stationary frameworks.

This growing recognition of the importance of variability has also exposed the fundamental limits of analytical approximations commonly used to bridge individual- and population-level dynamics. Spatial moment dynamics and coarse-grained density-based formulations provide tractable descriptions of spatial interactions, yet their accuracy deteriorates when local stochasticity, clustering, or strong nonlinearities dominate. In systems with Allee effects, moment closures can misestimate persistence thresholds and predict extinction, even when spatial structure enables survival \cite{Simpson}. More broadly, systematic comparisons show that neither moment-based nor density-based models reliably capture IBM outcomes across all parameter regimes, with each representation failing under conditions in which environmental variability and spatial correlations strongly shape population trajectories \cite{Surendran2025}. These limitations underscore the need for modeling approaches that fully resolve individual-level stochasticity and localized interactions. In this context, individual-based models (IBMs) provide a complementary bottom-up framework for studying vegetation dynamics with realistic variability. By explicitly representing birth, death, dispersal, and local interactions among discrete individuals, IBMs naturally incorporate demographic noise, finite-size effects, and localized positive feedbacks characteristic of Allee-type processes. Such density-dependent mechanisms—arising from microclimatic amelioration, shading, organic accumulation, or hydraulic facilitation—are well documented in plant systems and are particularly relevant in semi-arid environments, where the survival of small clusters often depends on stochastic sequences of favorable and unfavorable conditions \cite{Scanlon, Xu}. Therefore, IBMs offer a mechanistic approach to capturing the transient, heterogeneous, and noise-driven dynamics that increasingly define our understanding of patterned vegetation in drylands.

Here, we develop an individual-based model that builds directly on a recent deterministic formulation incorporating Allee effects and precipitation-dependent facilitation \cite{Gordillo23}, as well as on mechanistic insights into how the spatial structure modifies Allee thresholds \cite{Simpson}. Our model investigates how the interplay between local positive feedbacks and intermittent precipitation shapes vegetation structure and persistence in semi-arid landscapes. Facilitation intensity alternates between low and high values according to a stochastic switching process representing irregular rainfall events, capturing a defining feature of dryland hydroclimate: episodic pulses that temporarily enhance local survival and growth before conditions revert to competition-dominated dynamics. We show that this pulse-driven modulation of facilitation generates irregular, dynamically evolving clusters, which are in qualitative agreement with observations of vegetation mosaics in real drylands. We further quantify how the frequency and duration of favorable periods govern long-term persistence and show that the model's irregular spatial structures resemble those observed in natural settings.

\section{Theoretical framework}
\subsection{The local Allee effect}
The Allee effect describes a positive relationship between the per capita population growth rate and population density at low densities \cite{Courchamp}. Under conditions where birth rates exceed death rates, higher local density boosts survival and reproduction, driving population persistence and spatial expansion. In plant communities, this positive density dependence can promote aggregation and lead to the emergence of spatially clustered vegetation patterns \cite{Odum}. In our framework, the concepts of facilitation and facilitation influence are explicitly grounded in this local Allee effect. Following the formulation introduced in \cite{Simpson}, we derive a spatially dependent birth rate that captures positive density dependence through localized interactions among individuals.

\subsection{Modeling death and birth events in space}
We consider the dynamics of a plant population evolving within a square spatial domain of side length $L$, subject to periodic boundary conditions. Individuals are treated as discrete entities characterized by their spatial positions, and population dynamics proceed through stochastic birth and death events. The occurrence of birth and death for each individual depends on the spatial configuration of neighboring individuals, as detailed below. 

Individual death and birth rates are formulated as functions of competitive and facilitative interactions, which decay with inter-individual distance \cite{Simpson}. These interactions are described by Gaussian kernels $\psi_c$ and $\psi_f$, corresponding to competition and facilitation, respectively:
\begin{eqnarray}
    \psi_c (\xi)&= \gamma_c\exp{\left(-|\xi|^2/2\sigma_c^2\right)},\qquad \text{for competition influence,}\\
    \psi_f (\xi)&= \gamma_f\exp{\left(-|\xi|^2/2\sigma_f^2\right)},\qquad \text{for facilitation influence,}
\end{eqnarray}
where $\xi$ denotes the relative displacement between two individuals. Parameter values are reported in \cref{tab:parameter_values}.

Let $\mathbf{x}_k$ denote the position of the $k$th individual. At any given time, the cumulative competitive and facilitative influences experienced by the $n$th individual are given by
\begin{equation}
    X_n = \sum_{k\neq n} \psi_c(\bx_k-\bx_n) \quad \text{and} \quad Y_n= \sum_{k\neq n} \psi_f(\bx_k-\bx_n),
\end{equation}
respectively.

Given these interaction measures, the death and birth rates of individual $n$, denoted $D_n$ and $B_n$, are specified through the functions $F$ and $G$,
\begin{equation}
\begin{aligned} \label{eq: Allee rates}
    D_n &= F(X_n)= X_n^2+c\\
    B_n &= G(Y_n)= \rho + Y_n.
\end{aligned}
\end{equation}

These functional forms are derived from the classical per capita growth rate associated with a strong Allee effect \cite{Simpson}. The parameters $\rho$ and $c$ represent density-independent contributions to reproduction and death, respectively. To isolate the role of positive density dependence, we set $\rho = 0$ throughout, so that population growth is entirely driven by local interactions, i.e., the local Allee effect.

Following a birth event, offspring disperse spatially according to a normal distribution with zero mean and variance $\sigma^2$, reflecting short-range dispersal typical of many plant species. We consider that each reproductive event produces a single offspring.

\subsection{Modeling environmental stochastic effects}  
Environmental fluctuations in semiarid landscapes can alter both competition and facilitation. In particular, landscapes often experience periods of drought and irregular precipitation, that is, unpredictable variations in the timing, intensity, and distribution of rainfall events. These fluctuations in precipitation directly affect soil moisture levels, thereby influencing both competition and facilitation through shading and micro-habitat modification.

In our model, we consider precipitation events to occur at exponentially distributed times with a fixed rate $\lambda$ \cite{Rodriguez-Iturbe}, while their effects last for a fixed period of time $d$. The precipitation events are coupled to either an increase in facilitation or a decrease in competition, modifying the interaction kernels as,
\begin{eqnarray}\label{eq: facilitation}
    \psi_{f_+} (\xi)&= \gamma_{f_+}\exp{\left(-|\xi|^2/2\sigma_f^2\right)} \quad &\textrm{with} \quad \gamma_{f_+}>\gamma_f \\
    \psi_{c_-}(\xi)&= \gamma_{c_-}\exp{\left(-|\xi|^2/2\sigma_c^2\right)} \quad &\textrm{with} \quad \gamma_{c_-}<\gamma_c\ .
\end{eqnarray}

Thus, the dynamics of the system at any time $t$ is determined by the kernels in use. If precipitation is coupled to facilitation, $\psi_f$ is replaced by $\psi_{f_+}$ after a precipitation event and during $d$ time steps. On the other hand, when precipitation is coupled to competition, $\psi_c$ is replaced by $\psi_{c_-}$. 

We say that the system dynamics is in the low precipitation regime, $L$, if it is evolving with the kernel $\psi_f$ and $\psi_c$ and in the high precipitation regime, $H$, when it is evolving with any of the modified kernels, $\psi_{f_+}$ or $\psi_{c_-}$. We select a sufficiently long fixed simulation time to observe several alternations of the two dynamics, i.e., the different kernels. 

\subsection{Clustering detection: a pair-correlation function}
An interesting feature of the model is the irregular emergence and disappearance of vegetation clusters. Following \cite{Binder13}, we define a pair-correlation function (PCF), denoted by $C(|\xi|,t)$, where $|\xi|$ represents the separation distance between pairs of individuals. In short, this PCF counts the distances between individuals and is normalized so that $C(|\xi|,t)=1$ equals that of a random (uniform) distribution of individuals, i.e., complete spatial randomness (CSR) \cite{Binder11, Binder13, Diggle76}. In other words, for a fixed spatial distribution of individuals in a domain, the value $C(|\xi|,t)$, with $t$ fixed, compares the number of pairs of individuals at a specific distance $|\xi|$ from each other with the expected number of individuals if they were distributed uniformly in the domain considered. Thus, $C(|\xi|,t)=1$ at a specified distance $|\xi|$ suggests the absence of spatial correlation. Conversely, $C(|\xi|,t)>1$ suggests clustering (aggregation) at a distance $|\xi|$, whereas a value below 1 suggests spatial separation (segregation) \cite{Binder13}. Thus, this PCF function effectively quantifies the degree of spatial aggregation or segregation among individuals in our model.

\section{Results}

\subsection{Emergence and dynamics of vegetation clusters}

To explore the spatiotemporal dynamics generated by the model, we performed numerical simulations of a population evolving in a large $100 \times 100$ spatial domain with periodic boundary conditions and a starting random spatial distribution of $N=L^2=10^4$ individuals. The parameter values used in the simulation are summarized in \cref{tab:parameter_values}. These values were selected such that if the system evolved exclusively under the dry dynamics $L$ (low facilitation or high competition), the population would go extinct within the fixed simulation time (\siref{fig:example_no_switching_allee}). In contrast, if it evolved solely under the wet dynamics $H$ (high facilitation or low competition), the population would fluctuate around a stable positive equilibrium (\siref{fig:example_no_switching_allee}). 

Although it is common practice to model the onset of precipitation events with exponentially distributed times \cite{Rodriguez-Iturbe}, we first use the mean of such a distribution, $1/\lambda$, as the time between subsequent high-facilitation periods. We use this simplified scheme as a proxy for the stochastic sequence of wet and dry spells that characterizes semiarid climates \cite{Li2017WetDrySpellsChina}, but we later show that the same qualitative results are obtained by implementing stochastic precipitation events. Thus, we consider precipitation events to occur deterministically after a time $T=1/\lambda$ from the end of the effects of the previous precipitation event (which lasts for a given duration $d$), thereby allowing us to obtain a clearer picture of the model dynamics.

\begin{table}[H]
    \centering
    \caption{List of Parameters}
    \begin{tabular}{ccl}
        \hline
        \hline
        Parameter & Value & Description\\
        \hline
          $L$ & 100  & Spatial domain size ($L\times L$)\\
            $N$ & $10^4$ & Initial number of individuals\\
            $\lambda$ & $ \claudator{0,1}$ & Precipitation rate\\
            $d$ & $\claudator{0,50}$ & Time steps in state $S_h$\\
            $\sigma_c$ & 0.5 & Std. Dev. competition kernel\\
            $\sigma_f$ & 0.5 & Std. Dev. facilitation kernel\\
            $\sigma_d$ & 0.3 & Std. Dev. offspring dispersal kernel\\
            $\gamma_c$ & 1.05 & Competition strength\\
            $\gamma_f$ & 1.1 & Facilitation strength at low precipitation\\
            $\gamma_{f_+}$ & 2.42 & Facilitation strength at high precipitation\\
            $\gamma_{c_-}$ & 0.67 & Competition strength at high precipitation\\
            $c$ & 0.05 & Intrinsic death rate \\
        \hline \hline
    \end{tabular}
    \label{tab:parameter_values}
\end{table}

\begin{figure}[H]
    \centering
    \includegraphics[width=0.95\textwidth]{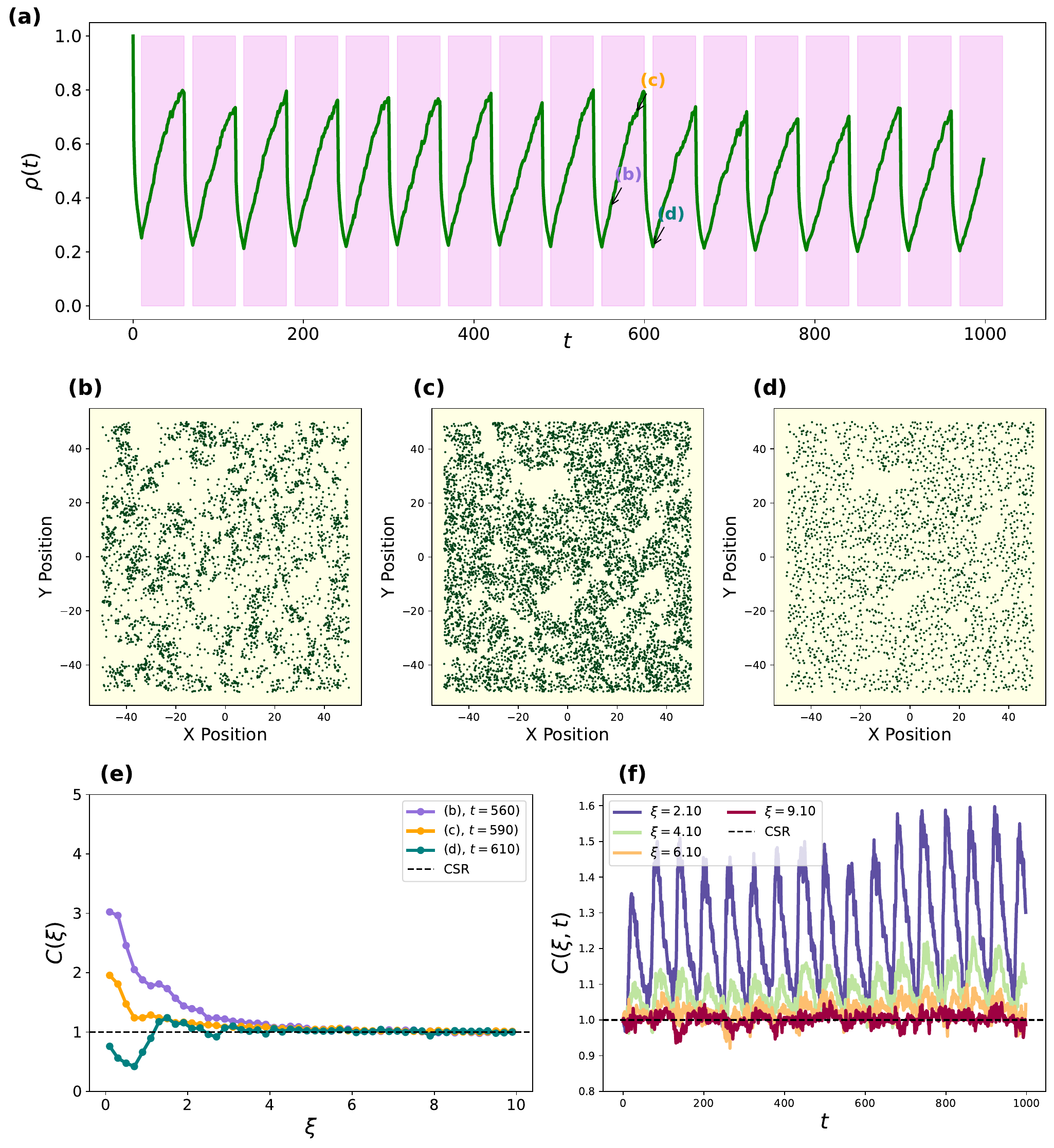}
    \caption{\textbf{Emergence and quantification of spatial clustering in the individual-based model (IBM) with precipitation-modulated competition.} (a) Trajectory of the normalized population density $\rho(t)$ over time. Pink shaded regions denote ``wet'' periods ($d=50$) where competition is reduced ($\psi_{c\_}$), mirroring rainfall pulses. (b–d) Spatial snapshots of individual plants at specific time points (indicated in panel a), showing the rapid formation of clusters during wet phases and their gradual thinning during dry intervals. (e) Pair-correlation function $C(\xi)$ calculated for snapshots (b)–(d); values $C(\xi) > 1$ at short distances indicate significant spatial aggregation (clustering) compared to complete spatial randomness (CSR). (f) Spatiotemporal evolution of the PCF for various distances $\xi$, highlighting how the local Allee effect maintains structural heterogeneity throughout the pulse-decay cycle.}
    \label{fig:allee_sim}
\end{figure}

The simulations reveal that intermittent modifications in either facilitation or competition, driven by precipitation events, generate a distinctive cycle of vegetation aggregation and decline. \cref{fig:allee_sim}(a) shows a representative trajectory of the normalized population density over a $100\times100$ spatial domain. Periods of decreased competition, preceded by precipitation events, are indicated by pink shading and last for a fixed duration $d=50$ time steps. It is clear that the alternation between low- and high-precipitation regimes produces oscillations in population density that mirror the temporal structure of rainfall.

To study the spatial dynamics, we computed the pair-correlation function (PCF) at all simulated times. The markers (b)–(d) indicate the time points at which snapshots of the spatial distributions were obtained and from which the PCFs shown in \cref{fig:allee_sim}(e) were computed. When a transition to the wet regime ($H$) occurs, such as in point (b), the balance between facilitation and competition increases, in this case by diminishing competition. Thus, through the local Allee effect, survival and reproduction are enhanced at short distances. This increase in positive interactions drives a rapid rise in population density and spontaneous formation of clusters, as evidenced by PCF values exceeding one at short spatial lags in points (b) and (c). These clusters represent locally reinforced patches in which individuals benefit from the presence of neighbors. Once the period of decreased competition ends, high competition dominates, and overall density declines. During this phase, individuals become spatially segregated, reducing local competition and stabilizing small surviving clusters (point (d)). As the population thins, natural mortality becomes the primary driver of change, and extinction risk peaks when clustering is at its lowest. If the population survives this bottleneck, the next precipitation event will renew facilitation, triggering a new cycle of cluster formation and decay (\cref{fig:allee_sim}(f), \textcolor{blue}{Supplementary Movie 1}). Of course, this clustering formation process driven by the Allee effect occurs over a limited spatial scale, precisely the characteristic length at which the interaction kernel becomes irrelevant. For this reason, the PCF at sufficiently high distances reduces to the complete spatial randomness limit (\cref{fig:allee_sim}(f)). Although offspring disperse over short distances and small clusters can form even in the absence of precipitation, it is the temporally intermittent amplification of the balance between facilitation and competition---mediated by the Allee effect---that promotes the emergence of larger, transient vegetation patches. This alternation between aggregation and decay underlies the irregular spatial mosaics observed in semiarid landscapes, where vegetation clusters appear and vanish in step with episodic rainfall.

The results shown in \cref{fig:allee_sim} were obtained by coupling precipitation to the competition kernel during the $H_p$ regime, but the same qualitative results are obtained when coupling precipitation to the facilitation kernel (\siref{fig:big_system_facilitation}, \textcolor{blue}{Supplementary Movie 2}). Similarly, the same qualitative results are obtained by implementing stochastic precipitation events(\siref{fig:big_system_facilitation_stochastic_precipitation,fig:big_system_competition_stochastic_precipitation}; \textcolor{blue}{Supplementary Movies 3} and \textcolor{blue}{4}).

\subsection{Absence of clustering without the Allee effect}

To isolate the specific role of the Allee effect in generating spatial clusters, we repeated the previous simulation by replacing the per capita birth and death rates in \cref{eq: Allee rates} with those of the logistic model,
\begin{equation}
\begin{aligned}
    D_n &= F(X_n)= X_n,\\
    B_n &= G(Y_n)=1,
\end{aligned}
\label{eq:logistic rates}
\end{equation}
thus removing the positive density dependence that characterizes the Allee effect. All other parameters were kept identical to those in \cref{fig:allee_sim}, with $\lambda=0.1$ and $d=50$. With these parameter values, the population would fluctuate around a different stable positive equilibrium in each of the precipitation regimes, $L_p$ and $H_p$ (\siref{fig:example_no_switching_logistic}). Because in the logistic model the birth rate is density independent, we coupled precipitation events to the competition kernel (decreasing competition). 

\begin{figure}[H]
    \centering
    \includegraphics[width=\textwidth]{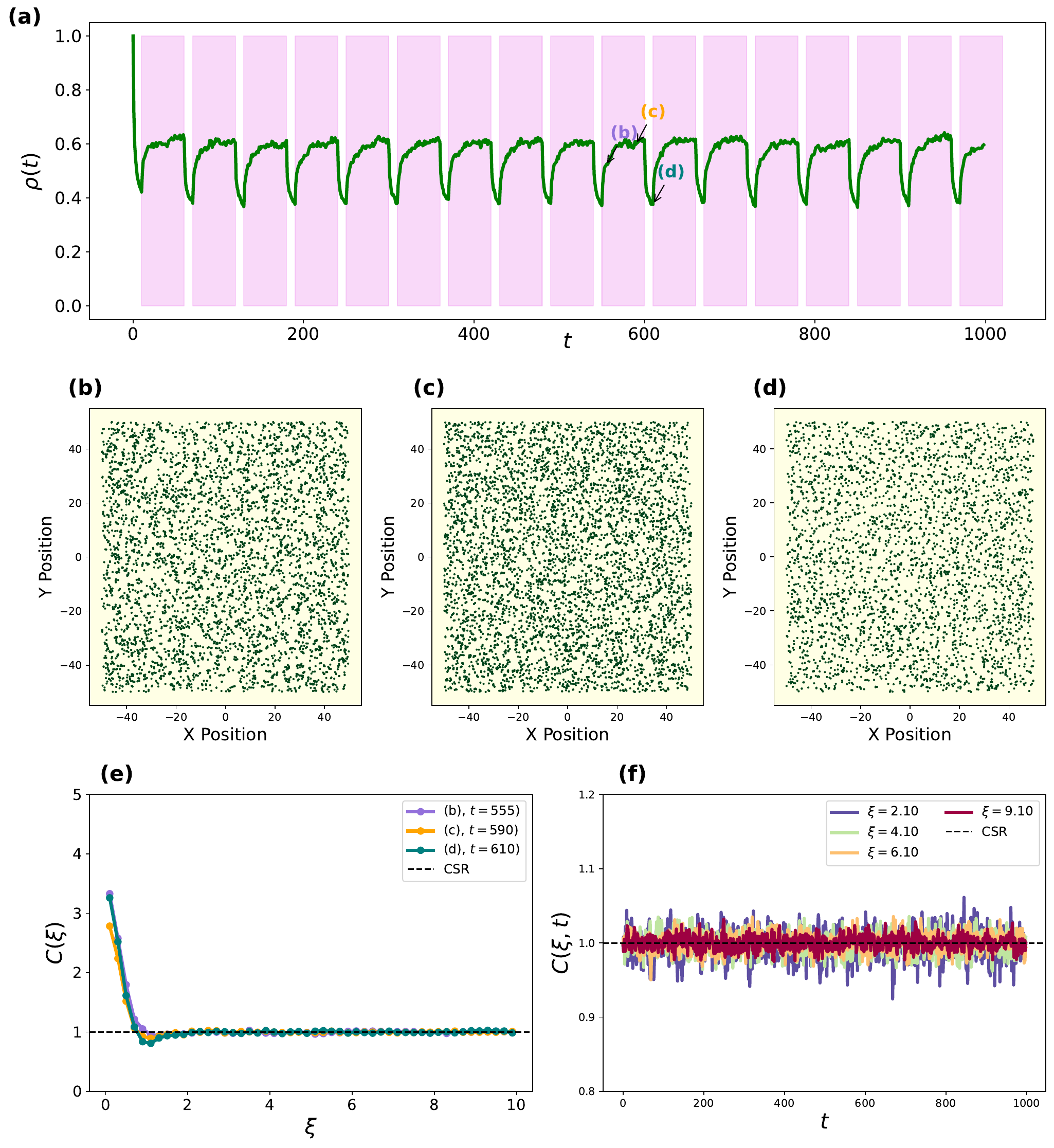}
    \caption{\textbf{Absence of large-scale clustering in the logistic model under intermittent precipitation.}. (a) Population density trajectory using a logistic model \cref{eq:logistic rates}. (b–d) Spatial snapshots showing a near-uniform distribution of individuals; while small, transient groups appear due to short-range dispersal, no large-scale clusters emerge. (e) The PCF remains near 1 across most distances, suggesting little spatial correlation. (f) Time-resolved PCF confirms that without positive density-dependent feedback, vegetation cannot self-organize into structured patterns despite environmental fluctuations.}
    \label{fig:logistic}
\end{figure}

Under these conditions, the population trajectories pivot around the steady states of the logistic model in each of the precipitation regimes (\cref{fig:logistic}(a), \siref{fig:example_no_switching_logistic}). The absence of a facilitative threshold prevents the sharp declines and recoveries observed in the Allee model. After each precipitation pulse, the population growth is smooth, and the density stabilizes without pronounced oscillations. The corresponding spatial configurations and pair-correlation functions reveal only short-range correlations (\cref{fig:logistic}(b-e)). Small clusters appear transiently because of the limited dispersal of offspring, but their size remains confined to the dispersal scale, and no larger aggregates emerge (\cref{fig:logistic}(e)). This phenomenology occurs at all distances and times (\cref{fig:logistic}(f), \textcolor{blue}{Supplementary Movie 5}). We also performed an additional simulation using $\gamma_c=1.8$ and $\gamma_{c_-}=0.6$, so that the difference between the steady states corresponding to high and low precipitation is enlarged (\siref{fig:example_no_switching_logistic_2}), obtaining the same results (\siref{fig:big_system_logistic_2}, \textcolor{blue}{Supplementary Movie 6}).

This contrast shows that demographic stochasticity and short-distance dispersal alone are insufficient to produce extended clustering. Instead, it is the interplay between the local balance of facilitation and competition, expressed here as the Allee effect, and its temporal modulation by precipitation events that generates large, transient vegetation patches. During wet periods, this balance amplifies local density increases, allowing neighboring individuals to reinforce each other’s survival and expansion. When conditions revert to dry phases, competition dominance over facilitation erodes these clusters, but their temporary formation imprints spatial heterogeneity on the landscape. 

In the absence of this density-dependent feedback, vegetation remains uniformly distributed and resilient to environmental fluctuations, yet is unable to self-organize into structured patterns. Therefore, the comparison confirms that the local Allee effect, intermittently enhanced by rainfall, is the fundamental mechanism driving the dynamic emergence and disappearance of vegetation clusters in our model.

\subsection{Survival probability under different precipitation regimes}

Next, we examined how the frequency and duration of precipitation events influence vegetation persistence. Specifically, we estimated the probability that the population remains extant after a sufficiently long simulation period under different combinations of $\lambda$ and $d$. To quantify these effects, we ran 100 independent realizations of the individual-based model for each pair $(d,\lambda)$, where $\lambda \in (0,1)$ and $d \in (2,50)$, and for a smaller system of size $L=50$. For each simulation, we recorded whether the population persisted until the end of the experiment and computed the fraction of surviving replicates as an empirical measure of survival probability. Because the system size is still big enough, fluctuations are small and different realizations are very similar to one another (\siref{fig:individual_realizations_allee_competition,fig:individual_realizations_allee_facilitation}).

\begin{figure}[H]
    \centering
   \includegraphics[width=0.9\linewidth]{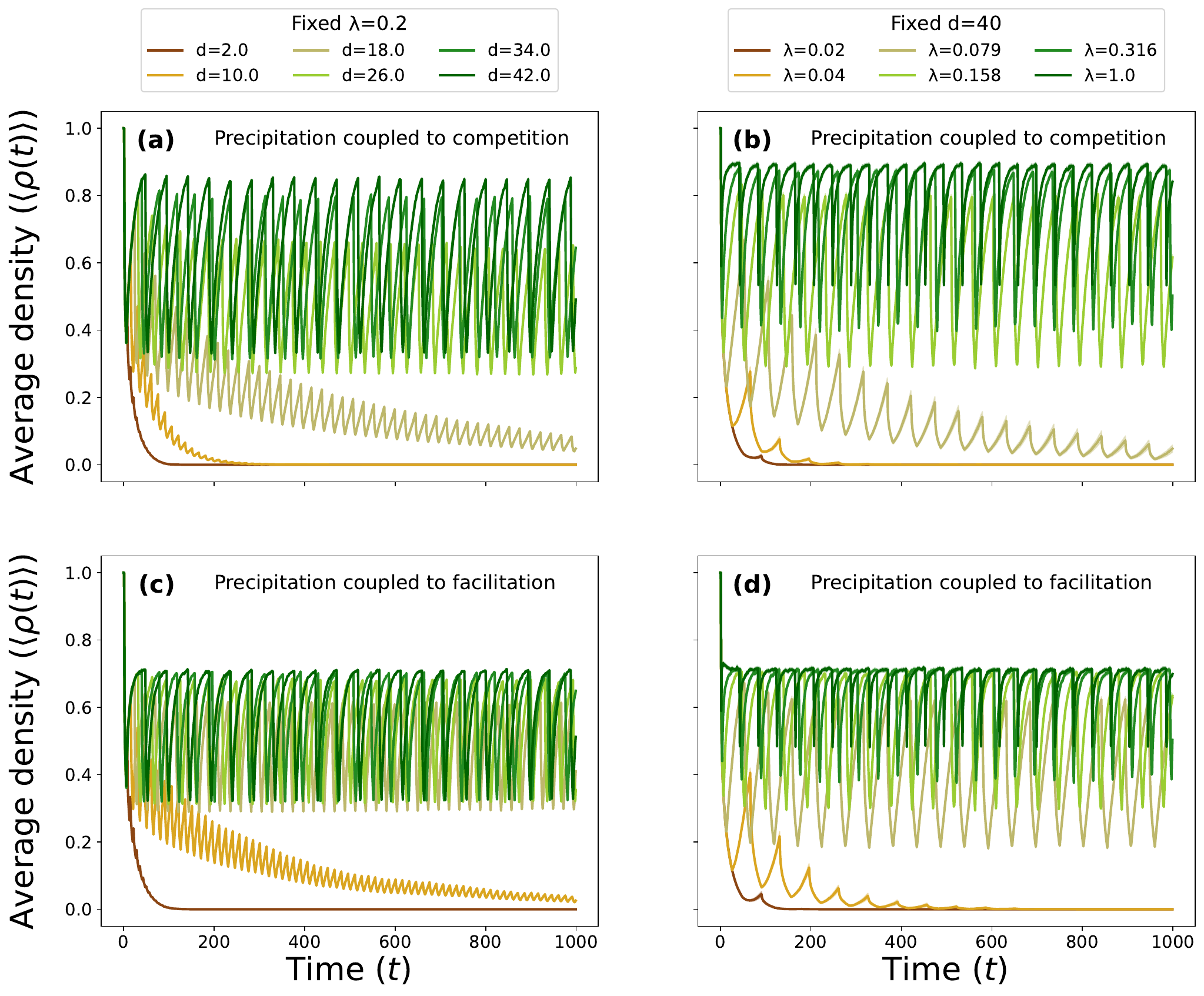}
    \caption{\textbf{Dynamics of average population density under different precipitation regimes, with precipitation coupled both to competition and facilitation.} (a,c) Trajectories for a fixed rainfall frequency ($\lambda = 0.1$) with varying pulse durations ($d$). Short durations lead to rapid extinction, while longer durations allow the Allee effect to amplify growth sufficiently to bridge dry intervals. (b,d) Trajectories for a fixed duration ($d = 40$) with varying frequencies ($\lambda$). Higher frequencies (shorter inter-pulse intervals) promote survival by preventing the population from dropping below the critical Allee threshold. The results are the average of $100$ realizations for a $50\cross50$ spatial domain, and the remaining parameter values used in the simulations are indicated in \cref{tab:parameter_values}. The shaded areas corresponding to 95\% confidence intervals are not visible. Individual realizations for some parameter values are shown in the \siref{fig:individual_realizations_allee_competition,fig:individual_realizations_allee_facilitation}.}
    \label{fig:sample paths}
\end{figure}

As expected, persistence strongly depends on the combination of the frequency and duration of favorable conditions. \cref{fig:sample paths} illustrates how these two parameters shape the temporal dynamics of the average population density. In \cref{fig:sample paths}(a,c), $\lambda$ is fixed at $0.2$, whereas the duration of the wet periods $d$ varies. Longer episodes of improved environmental conditions lead to pronounced increases in population density and delay extinction, as individuals experience extended phases of enhanced growth and survival. When $d$ is small, the facilitative impulse ends before the population recovers from prior declines, resulting in a rapid collapse. Conversely, when $d$ is sufficiently long, the Allee effect amplifies local growth, allowing clusters to expand and sustain a higher mean density throughout the subsequent dry period. 

\cref{fig:sample paths}(b,d) shows the complementary situation, with the duration fixed $d=40$ and the frequency of wet events $\lambda$ varied. Increasing $\lambda$ shortens the average waiting time between precipitation pulses, thereby providing more frequent opportunities for facilitation to counteract plant mortality. At low $\lambda$, long drought intervals separate consecutive favorable phases, and the population density decays to low values. At higher $\lambda$, successive facilitative episodes overlap in time, producing sustained oscillations around a positive mean density. 

\begin{figure}[H]
    \centering
\includegraphics[width=0.95\textwidth]{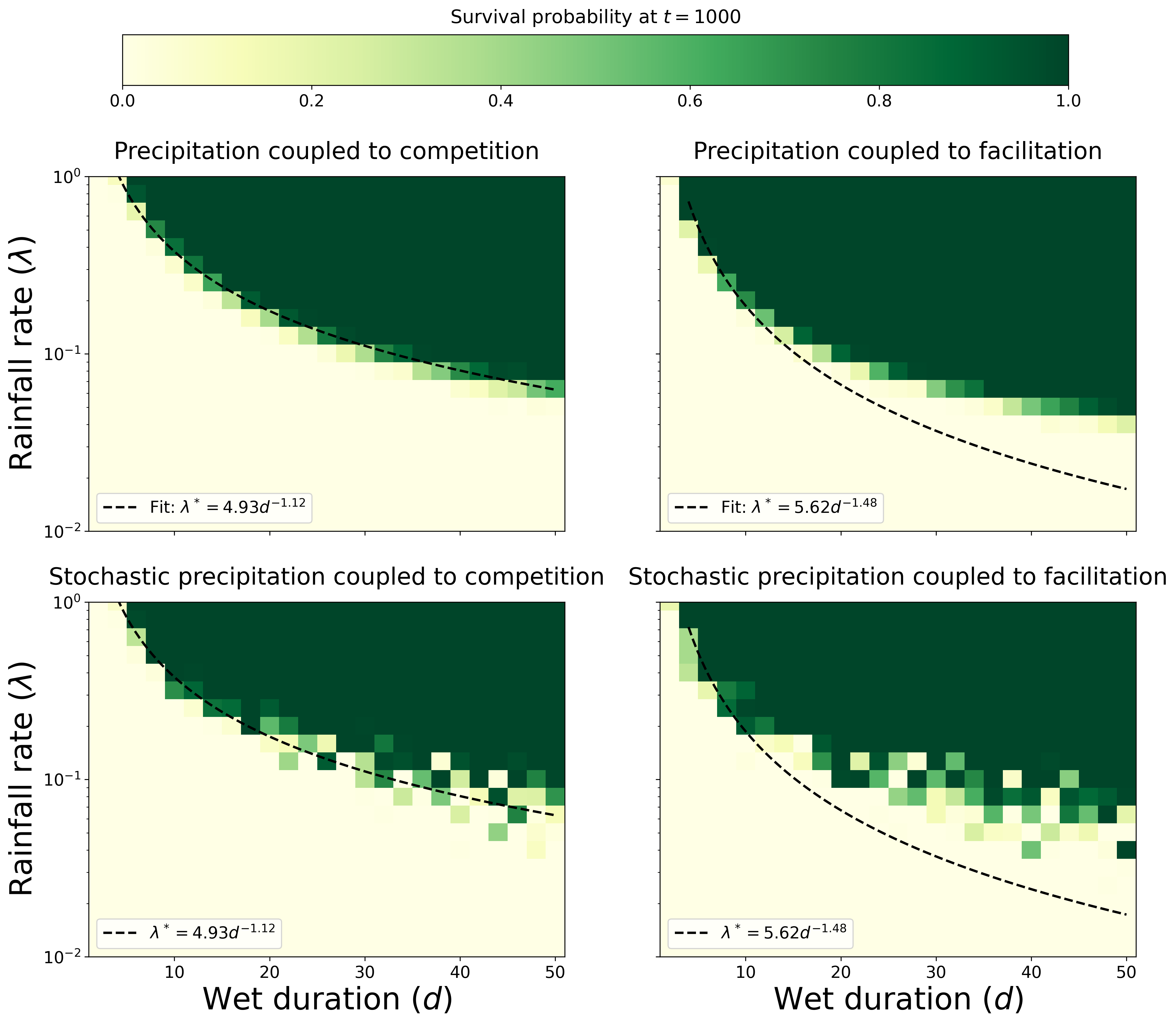}
    \caption{\textbf{Survival probability of vegetation as a function of precipitation frequency ($\lambda$) and duration ($d$).} The phase space reveals a sharp transition between certain extinction (yellow, low $\lambda$ and $d$) and reliable persistence (green, high $\lambda$ and $d$). The diagonal transition zone illustrates a compensatory mechanism: frequent short rainfall events can sustain the ecosystem as effectively as rare, long-lasting pulses. The black dashed line represents a power-law fit to the critical threshold, capturing the nonlinear boundary imposed by the spatial Allee effect. The phase diagram and the transition from extinction to persistence are largely insensitive to precipitation coupling-whether through increased facilitation or decreased competition-and remain robust to rainfall stochasticity. Other parameter values are indicated in \cref{tab:parameter_values}.}
    \label{fig:prob_survival}
\end{figure}

Together, these results reveal that persistence arises from a dynamic balance between the frequency and duration of the precipitation events, being these either coupled to competition or facilitation. Frequent short pulses can compensate for their brevity, whereas rare events can sustain the population if they are sufficiently long. In ecological terms, vegetation can withstand substantial variability in rainfall regimes as long as the cumulative exposure to facilitative conditions—controlled jointly by $\lambda$ and $d$—remains above a critical threshold required for recovery after dry periods.

\cref{fig:prob_survival} summarizes these results in the form of a phase diagram showing the probability of vegetation survival as a function of the precipitation parameters $\lambda$ and $d$. Survival is unlikely when both $\lambda$ and $d$ are small (yellow region), corresponding to rare, short precipitation events that fail to sustain facilitation long enough to offset mortality. In this regime, vegetation is repeatedly exposed to extended drought intervals during which the population drops below the Allee threshold, leading to extinction. As either $\lambda$ or $d$ increases, the cumulative exposure to favorable conditions also increases, and the population experiences more frequent or persistent bursts of growth. Consequently, the probability of survival increases sharply, forming a diagonal transition zone between collapse and persistence. Interestingly, the critical $d$ values at which the transition from extinction to persistence occurs depend on the rainfall rate following a power-law function (black dashed line).

This phase boundary reflects a compensatory relationship between the frequency and duration of precipitation events: frequent short episodes can maintain the population as effectively as sparse, long-lasting ones, provided that their combined effects exceed the critical facilitation threshold. The shape of the transition zone thus captures the intrinsic nonlinearity introduced by the Allee effect—vegetation persistence depends not on the mean rainfall alone, but on how rainfall variability modulates the temporal accumulation of facilitative interactions. Beyond this threshold, the population stabilizes through a succession of transient clusters that collectively maintain a positive density (green region), highlighting how intermittently enhanced facilitation can buffer semi-arid vegetation against extinction under fluctuating climatic conditions.

\subsubsection{Comparison with a mean-field model}

To assess the importance of spatial structure in determining vegetation persistence, we compared the outcomes of an individual-based model (IBM) with those of a mean-field (MF) approximation \cite{Gordillo23}. In the MF framework, all individuals interact equally with the entire population, effectively removing any dependence on spatial distance. This simplification allows us to isolate the effects of clustering and local facilitation, which are inherent to the IBM.

Let $u=u(t)$ denote the total population at time $t$. The cumulative competition and facilitation acting on a representative individual are obtained by integrating the respective kernels over the whole domain,
\begin{equation}
    X = u \!\int\! \psi_c(|\xi|)\,d\xi = 2\pi\sigma_c^2\gamma_cu, 
    \qquad 
    Y = u \!\int\! \psi_f(|\xi|)\,d\xi = 2\pi\sigma_f^2\gamma_fu,
\end{equation}
where we use a Gaussian form for the kernels. Substituting these expressions into the Allee-type birth and death rates \cref{eq: Allee rates} yields
\begin{equation}
\begin{aligned}\label{eq:mean-field rates}
    D &= \left(2\pi\sigma_c^2\gamma_cu\right)^2 + c,\\
    B &= 2\pi\sigma_f^2\gamma_fu.
\end{aligned}
\end{equation}
When facilitation is enhanced during wet phases, the corresponding birth rate becomes
\begin{equation}
\label{eq:mean-field switch}
    B_f = 2\pi\sigma_f^2\gamma_{f+}u.
\end{equation}
These expressions lead to two coupled differential equations that alternate stochastically between the dry ($L$) and wet ($H$) regimes, analogous to the switching process in the IBM:
\begin{eqnarray}\label{eq:mean-field}
    \dfrac{du}{dt} &= \Big(2\pi\big[\sigma_f^2\gamma_f - 2\pi\sigma_c^4(\gamma_c)^2u\big]u - c\Big)u, \\
    \dfrac{du}{dt} &= \Big(2\pi\big[\sigma_f^2\gamma_{f+} - 2\pi\sigma_c^4(\gamma_c)^2u\big]u - c\Big)u.
\end{eqnarray}

\begin{figure}[H]
    \centering   
    \includegraphics[width=0.8\textwidth]{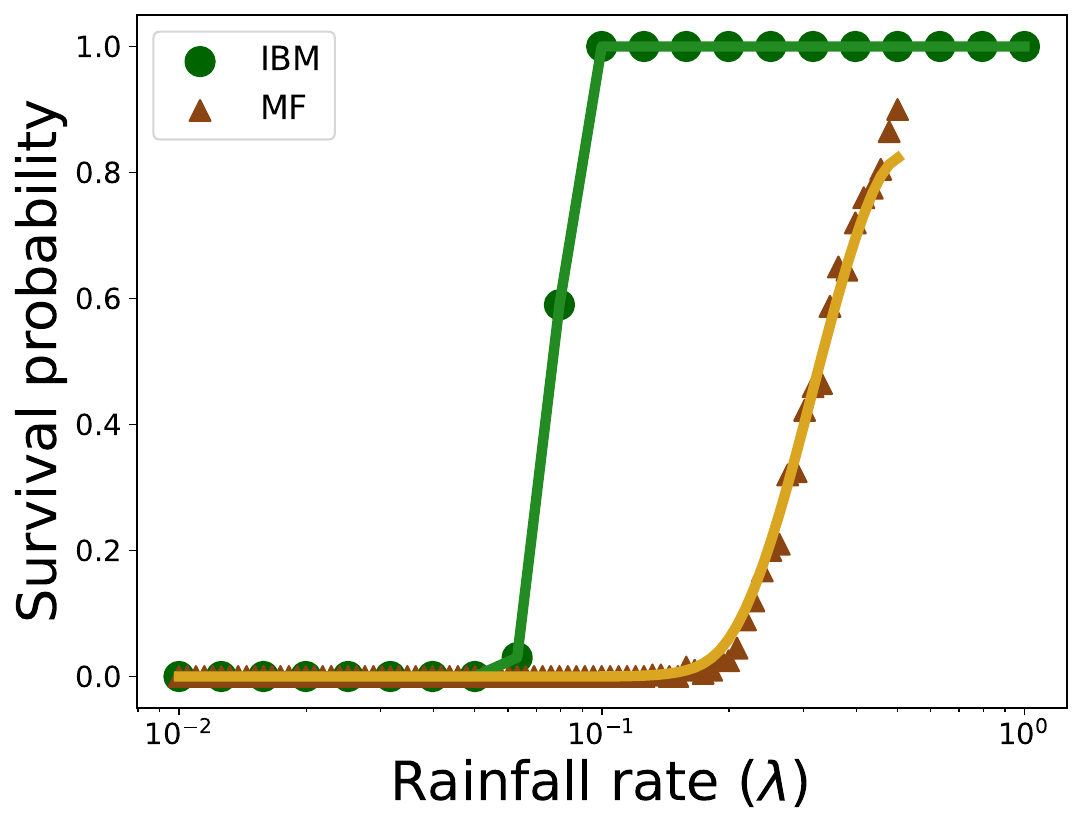}
    \caption{Probability of vegetation survival as a function of the precipitation rate, $\lambda$, for $d=25$. Average of 100 repetitions of the individual-based model with $L=50$ for each value of $\lambda$ (green circles) and a smooth fit (green curve). Average of 400 repetitions of the mean-field model for each value of $\lambda$ (brown triangles) and a smooth fit (brown curve). Other parameter values are indicated in \cref{tab:parameter_values}.}
    \label{fig:prob_survival mf}
\end{figure}

\cref{fig:prob_survival mf} compares the probability of population survival obtained from both models as a function of $\lambda$. Each data point represents the fraction of realizations that persisted until the end of the simulation, averaged over 400 replicates for the MF model and 100 for the IBM, using the same parameter values as in \cref{tab:parameter_values}. The MF model predicts markedly lower survival probabilities across all precipitation regimes. In the absence of spatial structure, the entire population experiences uniform competition and facilitation; thus, extinction occurs as soon as the global density falls below the Allee threshold. 

In contrast, the IBM retains localized clusters that act as refuges during dry periods. Within these patches, individuals benefit from short-range facilitation, which enhances local birth rates and buffers against stochastic mortality. When precipitation resumes, these surviving clusters serve as nuclei for recolonization, thereby increasing the population's overall persistence. Thus, the divergence between the two models quantifies the stabilizing effect of spatial heterogeneity: clustering not only amplifies the local action of the Allee effect but also enables recovery from near-extinction states that the MF approximation cannot capture.

This comparison underscores a central conclusion of this study: the resilience of vegetation in semiarid ecosystems is an emergent property of spatial interactions. Ignoring space, as in mean-field formulations, systematically underestimates persistence by neglecting the capacity of the spatial structure to maintain local facilitation and prevent global collapse.

\subsection{Comparison to observed vegetation clusters in semi-arid landscapes}

\begin{figure}[H]
    \centering
    \includegraphics[width=0.95\linewidth]{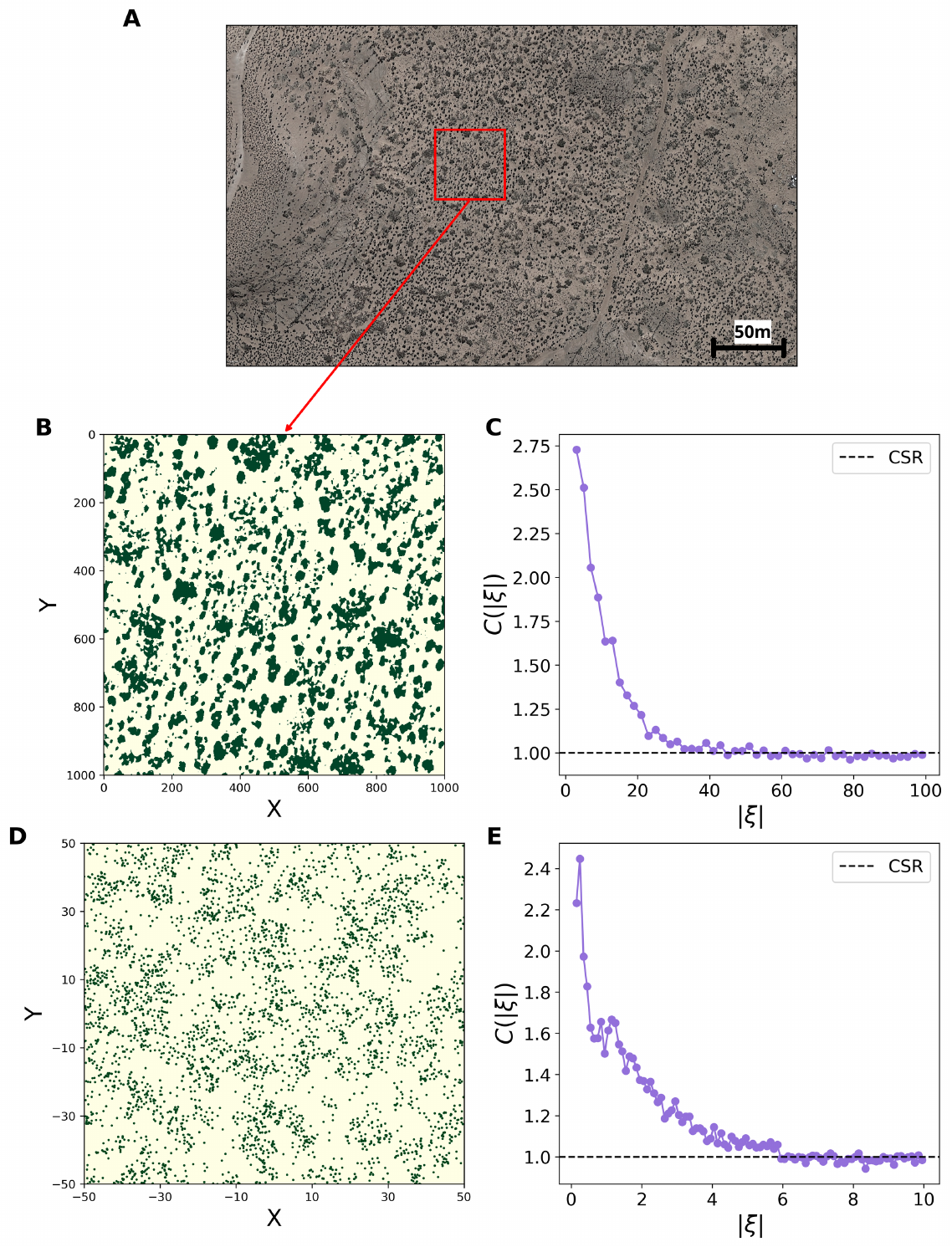}
    \caption{Comparison between simulated and observed vegetation patterns. (a,b) Zoomed subdomain of Horseshoe Bend (Arizona) displaying irregular vegetation clusters. (c) Pair-correlation function (PCFs) computed from the image. (d) Snapshot from a simulation of the individual-based model. (e) PCF of the simulation snapshot. Although the spatial scaled are markedly different, both simulated and observed patterns exhibit a similar aggregation pattern, illustrating that the stochastic alternation of local facilitation and competition in the model can reproduce realistic vegetation clustering in semiarid landscapes.}
    \label{fig:extended_space}
\end{figure}

To evaluate whether the model-generated spatial structures resemble those observed in natural ecosystems, we analyze a high-resolution aerial image of a semiarid region near Horseshoe Bend (Arizona) that was obtained from Google Earth  (\cref{fig:extended_space}(a)). Vegetation patches were automatically detected using a simple \textit{k}-means clustering algorithm applied to image intensity values  (\cref{fig:extended_space}(b)). From the binary map of vegetation presence, we computed the PCF using the same procedure as for the simulated data (\cref{fig:extended_space}(c)), enabling a direct comparison between empirical and model-generated spatial correlations (\cref{fig:extended_space}(d-e)).

This comparison reveals striking similarities between the two systems. Both the observed and simulated landscapes exhibit irregular vegetation clusters separated by sparsely vegetated gaps, with PCFs showing a clear short-range aggregation followed by a gradual decay to one, in contrast with the sharp declines of the PCF observed in the logistic model (\siref{fig:comparison_PCF_allee_logistic}). This pattern indicates a characteristic spatial scale of facilitation-driven clustering, beyond which plant distributions become uncorrelated. The close qualitative agreement between the simulated and observed PCFs suggests that the stochastic alternation of local facilitation and competition in the individual-based model captures the essential spatial mechanisms underlying vegetation self-organization in semiarid environments. Although simplified, the model reproduces the spatial heterogeneity typically observed in dryland ecosystems, supporting its potential as a tool for linking mechanistic theory with remote-sensing observations.

\section{Discussion}

Vegetation in semi-arid ecosystems is shaped by nonlinear interactions, demographic stochasticity, and strong climatic (intermittent) variability. Understanding how these ingredients jointly influence spatial organization and persistence remains a major challenge in the study of dryland resilience. Our results contribute to this effort by examining how the interplay between a local Allee effect and episodic hydrological forcing structures vegetation at the landscape scale. The emerging picture is one in which spatial patterns arise from a pulse-driven interplay between facilitation, competition, and stochastic fluctuations. These results highlight a general principle with increasing empirical support: vegetation resilience in drylands is shaped by the timing of favorable conditions as much as by their long-term mean \cite{Zhou2021,Smith2023}. When facilitative processes depend on rainfall pulses, the spacing and duration of these pulses can determine whether local populations cross Allee thresholds or collapse, suggesting that shifts in rainfall intermittency under climate change may have disproportionate effects on dryland stability—even if mean annual precipitation remains unchanged.

A critical insight from our comparative analysis is that spatial clustering functions as a demographic refuge, decoupling local persistence from global population trends. While the mean-field approximation predicts extinction whenever the global density falls below the Allee threshold, the individual-based model reveals that populations can persist globally even when the average density is vanishingly small, provided that the local density within clusters remains high. This role of aggregation in buffering demographic Allee effects is consistent with recent spatially explicit studies showing that aggregation can localize Allee constraints to the scale of clusters rather than the whole population \cite{Jorge2024}. In our framework, these clusters act as reservoirs of facilitation during dry dynamic phases, storing the population's reproductive potential until the next wet pulse triggers recolonization. This mechanism highlights that resilience in Allee-driven systems is not merely a property of total biomass but of the specific spatial covariance between individuals; the irregular, nonperiodic clusters are not just morphological outcomes of the dynamics but the functional structures that enable the population to persist through periods of environmental hostility that would otherwise be lethal.

The irregular spatial arrangements generated by our model resonate with a growing body of evidence showing that natural dryland vegetation rarely exhibits the highly periodic geometries predicted by classical reaction–diffusion frameworks \cite{Kästner2024,Pinto-Ramos2023,Yizhaq2014}. Field surveys and remote-sensing analyses consistently reveal disordered mosaics and temporal fluctuations in cover that depart markedly from the stationary patterns of autonomous PDE systems \cite{Deblauwe2008,LaPierre2016,Sanz2022}. This discrepancy has motivated a shift toward models that incorporate environmental variability, which has been shown to break periodicity and promote irregular cluster structures. Temporal fluctuations in water availability can drive intermittent percolation dynamics and broad patch-size distributions \cite{martin2020intermittent}, whereas impulsive reaction–diffusion formulations show that pulsed hydrological inputs alter the conditions for pattern onset and can suppress diffusion-driven instabilities \cite{Eigentler2020}. Spatial heterogeneity exerts similarly profound effects: local differences in mortality, soil-water redistribution, or competitive stress can generate pattern hysteresis, disordered aggregations, or smooth declines in biomass that contrast with the abrupt collapses predicted by deterministic models \cite{Pinto-Ramos2022,VillaMartin2015,Yizhaq2016,Yizhaq2014}. Our results are consistent with this broader literature and offer an individual-level mechanism (pulse-modulated facilitation) that complements these explanations. Furthermore, our results support the view that dryland vegetation patterns are fundamentally non-equilibrium structures that remain dynamically active, expanding, and contracting, and are shaped by fluctuating environmental forces. Simultaneously, our IBM formulation provides insight into how these dynamics unfold at the scale of discrete plants, clarifying the role of local Allee effects in mediating survival and aggregation under pulsed conditions. 

However, important limitations remain. Vegetation is represented by a single functional type with uniform mortality and facilitation parameters. Real drylands exhibit species-specific differences in rooting depth, drought tolerance, and facilitative capacity, which may mediate or amplify the effects of intermittent precipitation. Incorporating functional diversity can alter both the spatial and temporal structures of emergent patterns. In addition, the model assumes spatially homogeneous environmental forcing, except for temporal switching between favorable and unfavorable states. Heterogeneity in soil texture, microtopography, runoff pathways, or nutrient availability strongly influences pattern morphology and resilience \cite{Yizhaq2014,Pinto-Ramos2022}. Adding spatially structured variability could generate interactions between temporal pulses and landscape heterogeneity, potentially producing dynamics that are richer than those explored here. Furthermore, implementing rainfall intermittency via a stochastic switching process captures the pulsed character of dryland hydroclimate. However, it omits correlations and multi-scale structures present in real precipitation records \cite{Gherardi2019}. Moreover, the current model couples facilitation directly to precipitation events by assuming a given fixed duration, whereas in reality, this relationship is mediated by soil moisture dynamics, which introduce memory effects that could further modulate the duration of favorable conditions. More detailed representations, for example, through empirically derived rainfall sequences, could modify both the amplitude and timing of facilitation pulses and thus the conditions for persistence. Finally, although IBMs offer a natural way to incorporate individual-level variability, they are computationally demanding and remain analytically intractable beyond limited regimes. Recent studies have shown that moment-closure approximations and coarse-grained density models capture some aspects of the spatial structure but can fail in parameter regions where stochastic clustering dominates \cite{Simpson,Surendran2025}. Developing intermediate-scale models or improved closures that retain the essential effects of rainfall-driven facilitation pulses is an important avenue for future research.

From a theoretical perspective, developing intermediate-scale approximations that can accurately capture pulse-driven clustering dynamics remains an open mathematical challenge \cite{Surendran2025}. However, a promising extension of this work would be to use the model to assess the risk of desertification in semi-arid landscapes. This approach could combine the present individual-based framework with spatial information derived from remotely sensed imagery. High-resolution images acquired from drones, planes, or satellites would allow the extraction of vegetation patch distributions, from which spatial statistics, such as the pair-correlation function (PCF), can be computed, as illustrated here. In parallel, high-resolution precipitation data from products such as CHELSA \cite{chelsa-climatologies-2021} can be used to estimate the characteristic frequency of rainfall events, corresponding to the model parameter $\lambda$. By running the model across different kernel configurations and precipitation durations $d$, simulated spatial patterns could be generated and compared with the PCF to those observed in the imagery. This comparison would enable the model to be fitted to the specific vegetation dynamics of each site. Once calibrated, the model can be forced with precipitation data derived from climate change scenarios to estimate the temporal evolution of survival probability and, consequently, the likelihood of vegetation collapse under future climatic conditions. Such an approach would provide a mechanistic, spatially explicit framework for evaluating desertification risk by accounting for the interplay between vegetation spatial structure and precipitation variability.

\section*{Data and code availability}
The code and data used in this study are available in a \href{https://github.com/agimenezromero/Intermittent-precipitation-and-spatial-Allee-effects-drive-irregular-vegetation-patterns-in-semiarid}{GitHub} \cite{CODE} and a \href{https://doi.org/10.5281/zenodo.17974341}{Zenodo} \cite{Zenodo} repositories.

\section*{Acknowledgements}
We thank Ricardo Martínez-Garcia for his feedback on previous versions of this work. AGR and MAM acknowledge support by grants PID2021-123723OB-C22 (CYCLE), funded by the Spanish Ministry of Science and Innovation MICIU/AEI/10.13039/501100011033 and by ERDF, EU; and CEX2021-001164-M (María de Maeztu Program for Units of Excellence in R\&D) funded by MICIU/AEI/10.13039/501100011033. A.G.R. acknowledges financial support from grant JDC2024-053275-I, funded by MICIU/AEI/10.13039/501100011033 and FSE+; and from grant PID2024-156062OB-I00 (CHANGE-ME), funded by the Spanish Ministry of Science and Innovation MICIU/AEI/10.13039/501100011033.

\printbibliography

@article{Binder11,
  author = {Binder, B. J. and Landman, K. A.},
  title = {Quantifying evenly distributed states in exclusion and nonexclusion processes},
  journal = {Physical Review E},
  year = {2011},
  volume = {83},
  pages = {041914},
  doi={10.1103/PhysRevE.83.041914}
}

@article{Binder13,
  author = {Binder, B. J. and Simpson, M. J.},
  title = {Quantifying spatial structure in experimental observations and agent-based simulations using pair-correlation functions},
  journal = {Physical Review E},
  year = {2013},
  volume = {88},
  pages = {022705},
  doi={10.1103/PhysRevE.88.022705}
}

@article{Borgogno2009,
  author       = {Borgogno, Francesco and D'Odorico, Paolo and Laio, Francesco and Ridolfi, Luca},
  title        = {Mathematical models of vegetation pattern formation in ecohydrology},
  journal      = {Reviews of Geophysics},
  volume       = {47},
  number       = {RG1005},
  year         = {2009},
  doi          = {10.1029/2007RG000256},
  url          = {https://doi.org/10.1029/2007RG000256}
}

@book{Courchamp,
  author = {Courchamp, F. and Berec, L. and Gascoigne, J.},
  title = {Allee Effects in Ecology and Conservation},
  publisher = {Oxford University Press},
  address = {New York},
  year = {2008}
}

@article{Diggle76,
  author = {Diggle, P. J. and Besag, J. and Gleaves, J. T.},
  title = {Statistical analysis of spatial point patterns by means of distance methods},
  journal = {Biometrics},
  year = {1976},
  pages = {659--667},
  doi={10.2307/2529754}
}

@article{Gordillo23,
  author = {Gordillo, L. F. and Greenwood, P. E.},
  title = {Intermittent Precipitation-Dependent Interactions, Encompassing Allee Effect, May Yield Vegetation Patterns in a Transitional Parameter Range},
  journal = {Bulletin of Mathematical Biology},
  year = {2023},
  volume = {85},
  pages = {86},
  doi={10.1007/s11538-023-01191-y}
}

@article{Martinez-Garcia23,
  author = {Ricardo Martinez-Garcia and Ciro Cabal and Justin M. Calabrese and Emilio Hernández-García and Corina E. Tarnita and Cristóbal López and Juan A. Bonachela},
  title = {Integrating theory and experiments to link local mechanisms and ecosystem-level consequences of vegetation patterns in drylands},
  journal = {Chaos, Solitons \& Fractals},
  volume = {166},
  pages = {112881},
  year = {2023},
doi = {https://doi.org/10.1016/j.chaos.2022.112881}
}

@book{Odum,
  author = {Odum, E. P.},
  title = {Fundamentals of Ecology},
  edition = {2nd},
  publisher = {W.B. Saunders Company},
  address = {Philadelphia},
  year = {1959}
}

@article{Rietkerk04,
  author = {Rietkerk, M. and Dekker, S. C. and de Ruiter, P. C. and van de Koppel, J.},
  title = {Self-organized patchiness and catastrophic shifts in ecosystems},
  journal = {Science},
  year = {2004},
  volume = {305},
  number = {5692},
  pages = {1926--1929},
  doi = {https://doi.org/10.1126/science.1101867}
}

@article{Rietkerk21,
  author = {Max Rietkerk and Robbin Bastiaansen and Swarnendu Banerjee and Johan van de Koppel and
Mara Baudena and Arjen Doelman},
  title = {Evasion of tipping in complex systems through spatial pattern formation},
  journal = {Science},
  year = {2021},
  volume = {374},
  pages = {eabj0359},
  doi = {https://doi.org/10.1126/science.abj0359}
}

@book{Rodriguez-Iturbe,
  author = {Rodríguez-Iturbe, I. and Porporato, A.},
  title = {Ecohydrology of Water-Controlled Ecosystems},
  publisher = {Cambridge University Press},
  address = {Cambridge},
  year = {2004}
}

@article{Scanlon,
  author = {Scanlon, T. M. and Caylor, K. K. and Levin, S. A. and Rodríguez-Iturbe, I.},
  title = {Positive feedbacks promote power-law clustering of Kalahari vegetation},
  journal = {Nature},
  year = {2007},
  volume = {449},
  pages = {209--212},
  doi={10.1038/nature06060}
}

@article{Simpson,
  author = {Surendran, A. and Plank, M. J. and Simpson, M. J.},
  title = {Population dynamics with spatial structure and an Allee effect},
  journal = {Proceedings of the Royal Society A},
  year = {2020},
  volume = {476},
  pages = {20200501},
  doi={10.1098/rspa.2020.0501}
}

@article{Xu,
  author = {Chi Xu and Milena Holmgren and Egbert H. Van Nes and Fernando T. Maestre and Santiago Soliveres and Miguel Berdugo and Sonia Kéfi and Pablo A. Marquet and Sebastián Abades and Marten Scheffer},
  title = {Can we infer plant facilitation from remote sensing? A test across global drylands},
  journal = {Ecological Applications},
  year = {2015},
  volume = {25},
  number = {6},
  pages = {1456--1462},
  doi = { https://doi.org/10.1890/14-2358.1}
}

@article{Turing1952,
  author={Alan M. Turing},
  journal={Philosophical Transactions of the Royal Society B},
  voume={237},  
  pages={37-72},
  year={1952},
  doi={https://doi.org/10.1098%2Frstb.1952.0012}
}

@article{Lefever1997,
title = {On the origin of tiger bush},
journal = {Bulletin of Mathematical Biology},
volume = {59},
number = {2},
pages = {263-294},
year = {1997},
issn = {0092-8240},
doi = {https://doi.org/10.1016/S0092-8240(96)00072-9},
url = {https://www.sciencedirect.com/science/article/pii/S0092824096000729},
author = {R. Lefever and O. Lejeune}
}

@article{Rietkerk2008,
author = {Max Rietkerk and Johan {van de Koppel}},
title = {Regular pattern formation in real ecosystems},
journal = {Trends in Ecology \& Evolution},
volume = {23},
number = {3},
pages = {169-175},
year = {2008},
doi = {https://doi.org/10.1016/j.tree.2007.10.013},
url = {https://www.sciencedirect.com/science/article/pii/S0169534708000281}
}

@article{Klausmeier1999,
author = {Christopher A. Klausmeier},
title = {Regular and Irregular Patterns in Semiarid Vegetation},
journal = {Science},
volume = {284},
issue = {5421},
pages = {1826-1828},
year = {1999},
doi = {https://doi.org/10.1126/science.284.5421.1826},
}

@article{chelsa-climatologies-2021,
	 year = "2021",
	 publisher = "EnviDat",
      journal={EnviDat},
	 title = "Climatologies at high resolution for the earth’s land surface areas",
	 author = "Dirk Nikolaus Karger and Olaf Conrad and Jürgen Böhner and Tobias Kawohl and Holger Kreft and Rodrigo Wilber Soria-Auza and Niklaus E. Zimmermann and H. Peter Linder and Michael Kessler",
	 DOI = "10.16904/envidat.228",
howpublished={\url{https://doi.org/10.16904/envidat.228}}
}

@article{DeCoca2004,
author = {F. Camacho-De Coca and F. J. García-Haro and M. A. Gilabert and J. Meliá and},
title = {Vegetation cover seasonal changes assessment from TM imagery in a semi-arid landscape
},
journal = {International Journal of Remote Sensing},
volume = {25},
number = {17},
pages = {3451--3476},
year = {2004},
publisher = {Taylor \& Francis},
doi = {10.1080/01431160310001618761},
}

@article{Schmidt2000,
title = {Remote sensing of the seasonal variability of vegetation in a semi-arid environment},
journal = {Journal of Arid Environments},
volume = {45},
number = {1},
pages = {43-59},
year = {2000},
issn = {0140-1963},
doi = {https://doi.org/10.1006/jare.1999.0607},
url = {https://www.sciencedirect.com/science/article/pii/S0140196399906079},
author = {Heike Schmidt and Arnon Karnieli},
}

@Article{Sanz2022,
AUTHOR = {Sanz, Ernesto and Sotoca, Juan José Martín and Saa-Requejo, Antonio and Díaz-Ambrona, Carlos H. and Ruiz-Ramos, Margarita and Rodríguez, Alfredo and Tarquis, Ana M.},
TITLE = {Clustering Arid Rangelands Based on NDVI Annual Patterns and Their Persistence},
JOURNAL = {Remote Sensing},
VOLUME = {14},
YEAR = {2022},
NUMBER = {19},
ARTICLE-NUMBER = {4949},
URL = {https://www.mdpi.com/2072-4292/14/19/4949},
ISSN = {2072-4292},
DOI = {10.3390/rs14194949}
}

@book{assessment2005ecosystems,
  title={Ecosystems and human well-being: wWtlands and water},
  author={{Millennium Ecosystem Assessment}},
  year={2005},
  publisher={World Resources Institute},
howpublished = "\url{https://www.millenniumassessment.org/documents/document.358.aspx.pdf}",
url={https://www.millenniumassessment.org/documents/document.358.aspx.pdf}
}

@Article{Huang2016,
author={Huang, Jianping
and Yu, Haipeng
and Guan, Xiaodan
and Wang, Guoyin
and Guo, Ruixia},
title={Accelerated dryland expansion under climate change},
journal={Nature Climate Change},
year={2016},
day={01},
volume={6},
number={2},
pages={166-171},
issn={1758-6798},
doi={10.1038/nclimate2837},
url={https://doi.org/10.1038/nclimate2837}
}

@article{Deblauwe2008,
author = {Deblauwe, Vincent and Barbier, Nicolas and Couteron, Pierre and Lejeune, Olivier and Bogaert, Jan},
title = {The global biogeography of semi-arid periodic vegetation patterns},
journal = {Global Ecology and Biogeography},
volume = {17},
number = {6},
pages = {715-723},
doi = {https://doi.org/10.1111/j.1466-8238.2008.00413.x},
url = {https://onlinelibrary.wiley.com/doi/abs/10.1111/j.1466-8238.2008.00413.x},
eprint = {https://onlinelibrary.wiley.com/doi/pdf/10.1111/j.1466-8238.2008.00413.x},
year = {2008}
}

@article{Zhenpeng2023,
author = {Zhenpeng Ge },
title = {The hidden order of Turing patterns in arid and semi‐arid vegetation ecosystems},
journal = {Proceedings of the National Academy of Sciences},
volume = {120},
number = {42},
pages = {e2306514120},
year = {2023},
doi = {10.1073/pnas.2306514120},
URL = {https://www.pnas.org/doi/abs/10.1073/pnas.2306514120},
eprint = {https://www.pnas.org/doi/pdf/10.1073/pnas.2306514120},
}

@article{Rietkerk2002,
author = {Rietkerk, Max and Boerlijst, Maarten C. and van Langevelde, Frank and HilleRisLambers, Reinier and de Koppel, Johan van and Kumar, Lalit and Prins, Herbert H. T. and de Roos, Andr\'{e} M.},
title = {Self‐Organization of Vegetation in Arid Ecosystems.},
journal = {The American Naturalist},
volume = {160},
number = {4},
pages = {524-530},
year = {2002},
doi = {10.1086/342078},
anote ={PMID: 18707527},
URL={https://doi.org/10.1086/342078},
eprint={https://doi.org/10.1086/342078}
}

@Article{Kéfi2007,
author={K{\'e}fi, Sonia
and Rietkerk, Max
and Alados, Concepci{\'o}n L.
and Pueyo, Yolanda
and Papanastasis, Vasilios P.
and ElAich, Ahmed
and de Ruiter, Peter C.},
title={Spatial vegetation patterns and imminent desertification in Mediterranean arid ecosystems},
journal={Nature},
year={2007},
day={01},
volume={449},
number={7159},
pages={213-217},
issn={1476-4687},
doi={10.1038/nature06111},
url={https://doi.org/10.1038/nature06111}
}

@article{Kefi2014,
    doi = {10.1371/journal.pone.0092097},
    author = {Kéfi, Sonia AND Guttal, Vishwesha AND Brock, William A. AND Carpenter, Stephen R. AND Ellison, Aaron M. AND Livina, Valerie N. AND Seekell, David A. AND Scheffer, Marten AND van Nes, Egbert H. AND Dakos, Vasilis},
    journal = {PLoS ONE},
    publisher = {Public Library of Science},
    title = {Early Warning Signals of Ecological Transitions: Methods for Spatial Patterns},
    year = {2014},
    month = {03},
    volume = {9},
    url = {https://doi.org/10.1371/journal.pone.0092097},
    pages = {1-13},
    number = {3},
}

@article{Scheffer2012,
author = {Marten Scheffer  and Stephen R. Carpenter  and Timothy M. Lenton  and Jordi Bascompte  and William Brock  and Vasilis Dakos  and Johan van de Koppel  and Ingrid A. van de Leemput  and Simon A. Levin  and Egbert H. van Nes  and Mercedes Pascual  and John Vandermeer },
title = {Anticipating Critical Transitions},
journal = {Science},
volume = {338},
number = {6105},
pages = {344-348},
year = {2012},
doi = {10.1126/science.1225244},
URL = {https://www.science.org/doi/abs/10.1126/science.1225244},
eprint = {https://www.science.org/doi/pdf/10.1126/science.1225244}}

@book{Whitford2019,
  title={Ecology of desert systems},
  author={Whitford, Walter G and Duval, Benjamin D},
  year={2019},
  publisher={Academic Press}
}

@article{Gilad2004,
  title = {Ecosystem Engineers: From Pattern Formation to Habitat Creation},
  author = {Gilad, E. and von Hardenberg, J. and Provenzale, A. and Shachak, M. and Meron, E.},
  journal = {Physical Review Letters},
  volume = {93},
  issue = {9},
  pages = {098105},
  numpages = {4},
  year = {2004},

  publisher = {American Physical Society},
  doi = {10.1103/PhysRevLett.93.098105},
  url = {https://link.aps.org/doi/10.1103/PhysRevLett.93.098105}
}

@article{Gilad2007,
title = {Dynamics and spatial organization of plant communities in water-limited systems},
journal = {Theoretical Population Biology},
volume = {72},
number = {2},
pages = {214-230},
year = {2007},
issn = {0040-5809},
doi = {https://doi.org/10.1016/j.tpb.2007.05.002},
url = {https://www.sciencedirect.com/science/article/pii/S0040580907000603},
author = {E. Gilad and M. Shachak and E. Meron}
}

@article{Martínez-García2013,
author = {Martínez-García, Ricardo and Calabrese, Justin M. and Hernández-García, Emilio and López, Cristóbal},
title = {Vegetation pattern formation in semiarid systems without facilitative mechanisms},
journal = {Geophysical Research Letters},
volume = {40},
number = {23},
pages = {6143-6147},
doi = {https://doi.org/10.1002/2013GL058797},
url = {https://agupubs.onlinelibrary.wiley.com/doi/abs/10.1002/2013GL058797},
eprint = {https://agupubs.onlinelibrary.wiley.com/doi/pdf/10.1002/2013GL058797},
year = {2013}
}

@article{Li2017WetDrySpellsChina,  
  author = {Z. Li and Y. Li and X. Shi and J. Li},  
  title = {The characteristics of wet and dry spells for the diverse climate in China},  
  journal = {Global and Planetary Change},  
  volume = {149},  
  pages = {14--19},  
  year = {2017},  
  doi = {10.1016/j.gloplacha.2016.12.015}  
}

@article{martin2020intermittent,
  title={Intermittent percolation and the scale-free distribution of vegetation clusters},
  author={Martín, Paula Villa and Domínguez-García, Virginia and Muñoz, Miguel A},
  journal={New Journal of Physics},
  volume={22},
  number={8},
  pages={083014},
  year={2020},
  doi={10.1088/1367-2630/ab9f6e},
  publisher={IOP Publishing}
}

@article{Eigentler2020,
title = {Effects of precipitation intermittency on vegetation patterns in semi-arid landscapes},
journal = {Physica D: Nonlinear Phenomena},
volume = {405},
pages = {132396},
year = {2020},
issn = {0167-2789},
doi = {https://doi.org/10.1016/j.physd.2020.132396},
url = {https://www.sciencedirect.com/science/article/pii/S0167278919305238},
author = {L. Eigentler and J.A. Sherratt},
}

@article{Yizhaq2014,
author = {Yizhaq, H. and Sela, S. and Svoray, T. and Assouline, S. and Bel, G.},
title = {Effects of heterogeneous soil-water diffusivity on vegetation pattern formation},
journal = {Water Resources Research},
volume = {50},
number = {7},
pages = {5743-5758},
doi = {https://doi.org/10.1002/2014WR015362},
url = {https://agupubs.onlinelibrary.wiley.com/doi/abs/10.1002/2014WR015362},
eprint = {https://agupubs.onlinelibrary.wiley.com/doi/pdf/10.1002/2014WR015362},
year = {2014}
}

@article{Pinto-Ramos2023,
author = {D. Pinto-Ramos  and M. G. Clerc  and M. Tlidi },
title = {Topological defects law for migrating banded vegetation patterns in arid climates},
journal = {Science Advances},
volume = {9},
number = {31},
pages = {eadf6620},
year = {2023},
doi = {10.1126/sciadv.adf6620},
URL = {https://www.science.org/doi/abs/10.1126/sciadv.adf6620},
eprint = {https://www.science.org/doi/pdf/10.1126/sciadv.adf6620},
}

@article{Kästner2024,
title = {A scale-invariant method for quantifying the regularity of environmental spatial patterns},
journal = {Ecological Complexity},
volume = {60},
pages = {101104},
year = {2024},
issn = {1476-945X},
doi = {https://doi.org/10.1016/j.ecocom.2024.101104},
url = {https://www.sciencedirect.com/science/article/pii/S1476945X24000321},
author = {Karl Kästner and Roeland C. {van de Vijsel} and Daniel Caviedes-Voullième and Christoph Hinz},
}

@article{Pinto-Ramos2022,
title = {Vegetation covers phase separation in inhomogeneous environments},
journal = {Chaos, Solitons \& Fractals},
volume = {163},
pages = {112518},
year = {2022},
issn = {0960-0779},
doi = {https://doi.org/10.1016/j.chaos.2022.112518},
url = {https://www.sciencedirect.com/science/article/pii/S0960077922007184},
author = {D. Pinto-Ramos and S. Echeverría-Alar and M.G. Clerc and M. Tlidi},
}

@article{Odorico2006,
author = {D'Odorico, Paolo and Laio, Francesco and Ridolfi, Luca},
title = {Vegetation patterns induced by random climate fluctuations},
journal = {Geophysical Research Letters},
volume = {33},
number = {19},
pages = {},
doi = {https://doi.org/10.1029/2006GL027499},
url = {https://agupubs.onlinelibrary.wiley.com/doi/abs/10.1029/2006GL027499},
eprint = {https://agupubs.onlinelibrary.wiley.com/doi/pdf/10.1029/2006GL027499},
year = {2006}
}

@article{Yizhaq2016,
doi = {10.1088/1367-2630/18/2/023004},
url = {https://doi.org/10.1088/1367-2630/18/2/023004},
year = {2016},
month = {01},
publisher = {IOP Publishing},
volume = {18},
number = {2},
pages = {023004},
author = {Yizhaq, Hezi and Bel, Golan},
title = {Effects of quenched disorder on critical transitions in pattern-forming systems},
journal = {New Journal of Physics},
}

@article{VillaMartin2015,
author = {Paula Villa Martín  and Juan A. Bonachela  and Simon A. Levin  and Miguel A. Muñoz },
title = {Eluding catastrophic shifts},
journal = {Proceedings of the National Academy of Sciences},
volume = {112},
number = {15},
pages = {E1828-E1836},
year = {2015},
doi = {10.1073/pnas.1414708112},
URL = {https://www.pnas.org/doi/abs/10.1073/pnas.1414708112},
eprint = {https://www.pnas.org/doi/pdf/10.1073/pnas.1414708112},
}

@article{LaPierre2016,
author={La Pierre, Kimberly J.
and Blumenthal, Dana M.
and Brown, Cynthia S.
and Klein, Julia A.
and Smith, Melinda D.},
title={Drivers of Variation in Aboveground Net Primary Productivity and Plant Community Composition Differ Across a Broad Precipitation Gradient},
journal={Ecosystems},
year={2016},
day={01},
volume={19},
number={3},
pages={521-533},
issn={1435-0629},
doi={10.1007/s10021-015-9949-7},
url={https://doi.org/10.1007/s10021-015-9949-7}
}

@article{Gherardi2019,
author = {Gherardi, Laureano A. and Sala, Osvaldo E.},
title = {Effect of interannual precipitation variability on dryland productivity: A global synthesis},
journal = {Global Change Biology},
volume = {25},
number = {1},
pages = {269-276},
doi = {https://doi.org/10.1111/gcb.14480},
url = {https://onlinelibrary.wiley.com/doi/abs/10.1111/gcb.14480},
eprint = {https://onlinelibrary.wiley.com/doi/pdf/10.1111/gcb.14480},
year = {2019}
}

@article{Szeles2025,
title = {Comparative analysis of rainfall event characteristics and rainfall erosivity between two experimental plots in Austria and Slovenia},
journal = {Journal of Hydrology: Regional Studies},
volume = {59},
pages = {102353},
year = {2025},
issn = {2214-5818},
doi = {https://doi.org/10.1016/j.ejrh.2025.102353},
url = {https://www.sciencedirect.com/science/article/pii/S2214581825001788},
author = {Borbala Szeles and Juraj Parajka and Mojca Šraj and Günter Blöschl and Dušan Marjanović and Nejc Bezak and Klaudija Lebar and Andrej Vidmar and Peter Strauss and Carmen Krammer and Elmar Schmaltz and Patrick Hogan and Gerhard Rab and Katarina Zabret},
}

@article{Yizhaq2017,
title = {Geodiversity increases ecosystem durability to prolonged droughts},
journal = {Ecological Complexity},
volume = {31},
pages = {96-103},
year = {2017},
issn = {1476-945X},
doi = {https://doi.org/10.1016/j.ecocom.2017.06.002},
url = {https://www.sciencedirect.com/science/article/pii/S1476945X17300478},
author = {Hezi Yizhaq and Ilan Stavi and Moshe Shachak and Golan Bel},
}

@article{Pinto-Ramos2025,
author = {Pinto-Ramos, David and Clerc, Marcel Gabriel and Makhoute, Abdelkader and Tlidi, Mustapha},
title = {Aperiodic Clustered and Periodic Hexagonal Vegetation Spot Arrays Explained by Inhomogeneous Environments and Climate Trends in Arid Ecosystems},
journal = {Geophysical Research Letters},
volume = {52},
number = {21},
pages = {e2025GL118462},
doi = {https://doi.org/10.1029/2025GL118462},
url = {https://agupubs.onlinelibrary.wiley.com/doi/abs/10.1029/2025GL118462},
eprint = {https://agupubs.onlinelibrary.wiley.com/doi/pdf/10.1029/2025GL118462},
note = {e2025GL118462 2025GL118462},
year = {2025}
}

@article{Surendran2025,
title = {Spatial moment dynamics and biomass density equations provide complementary, yet limited, descriptions of pattern formation in individual-based simulations},
journal = {Physica D: Nonlinear Phenomena},
volume = {477},
pages = {134703},
year = {2025},
issn = {0167-2789},
doi = {https://doi.org/10.1016/j.physd.2025.134703},
url = {https://www.sciencedirect.com/science/article/pii/S0167278925001800},
author = {Anudeep Surendran and David Pinto-Ramos and Rafael Menezes and Ricardo Martinez-Garcia},
}

@article{Smith2023,
author={Smith, Taylor
and Boers, Niklas},
title={Global vegetation resilience linked to water availability and variability},
journal={Nature Communications},
year={2023},
month={01},
day={30},
volume={14},
number={1},
pages={498},
issn={2041-1723},
doi={10.1038/s41467-023-36207-7},
url={https://doi.org/10.1038/s41467-023-36207-7}
}

@article{Zhou2021,
title = {Characterizing vegetation response to rainfall at multiple temporal scales in the Sahel-Sudano-Guinean region using transfer function analysis},
journal = {Remote Sensing of Environment},
volume = {252},
pages = {112108},
year = {2021},
issn = {0034-4257},
doi = {https://doi.org/10.1016/j.rse.2020.112108},
url = {https://www.sciencedirect.com/science/article/pii/S0034425720304818},
author = {Jie Zhou and Li Jia and Massimo Menenti and Mattijn {van Hoek} and Jing Lu and Chaolei Zheng and Hao Wu and Xiaotian Yuan},
}

@misc{CODE,
author={Giménez-Romero, {\`A}lex},
year={2026},
howpublished={\url{https://github.com/agimenezromero/IBM-vegetation-dynamics-with-Allee-effect-and-intermittent-precipitation}}
}

@article{Zenodo,
  author    = {Giménez-Romero, Àlex},
  title     = {{Intermittent precipitation and spatial Allee effects drive irregular vegetation patterns in semiarid ecosystems}},
  month     = 02,
  year      = 2026,
  publisher = {Zenodo},
  doi       = {10.5281/zenodo.18470198},
  url       = {https://doi.org/10.5281/zenodo.18470198}
}

@article{Jorge2024,
    author = {Jorge, Daniel C. P. and Martinez-Garcia, Ricardo},
    title = {Demographic effects of aggregation in the presence of a component Allee effect},
    journal = {Journal of The Royal Society Interface},
    volume = {21},
    number = {215},
    pages = {20240042},
    year = {2024},
    month = {06},
    issn = {1742-5689},
    doi = {10.1098/rsif.2024.0042},
    url = {https://doi.org/10.1098/rsif.2024.0042},
    eprint = {https://royalsocietypublishing.org/rsif/article-pdf/doi/10.1098/rsif.2024.0042/929628/rsif.2024.0042.pdf},
}

@article{Collins2014,
   author = "Collins, S.L. and Belnap, J. and Grimm, N.B. and Rudgers, J.A. and Dahm, C.N. and D&apos;Odorico, P. and Litvak, M. and Natvig, D.O. and Peters, D.C. and Pockman, W.T. and Sinsabaugh, R.L. and Wolf, B.O.",
   title = "A Multiscale, Hierarchical Model of Pulse Dynamics in Arid-Land Ecosystems", 
   journal= "Annual Review of Ecology, Evolution, and Systematics",
   year = "2014",
   volume = "45",
   number = "Volume 45, 2014",
   pages = "397-419",
   doi = "https://doi.org/10.1146/annurev-ecolsys-120213-091650",
   url = "https://www.annualreviews.org/content/journals/10.1146/annurev-ecolsys-120213-091650",
   publisher = "Annual Reviews",
   issn = "1545-2069",
   type = "Journal Article",
}

@article{Kefi2024,
author = {Sonia Kéfi  and Alexandre Génin  and Angeles Garcia-Mayor  and Emilio Guirado  and Juliano S. Cabra
l  and Miguel Berdugo  and Josquin Guerber  and Ricard Solé  and Fernando T. Maestre },
title = {Self-organization as a mechanism of resilience in dryland ecosystems},
journal = {Proceedings of the National Academy of Sciences},
volume = {121},
number = {6},
pages = {e2305153121},
year = {2024},
doi = {10.1073/pnas.2305153121},
URL = {https://www.pnas.org/doi/abs/10.1073/pnas.2305153121},
eprint = {https://www.pnas.org/doi/pdf/10.1073/pnas.2305153121},
}

\end{document}


\noindent{\LARGE{\textbf{Supplementary Information for}}} 

{\let\newpage\relax\maketitle}

    \begin{figure}[H]
        \centering
        \includegraphics[width=\linewidth]{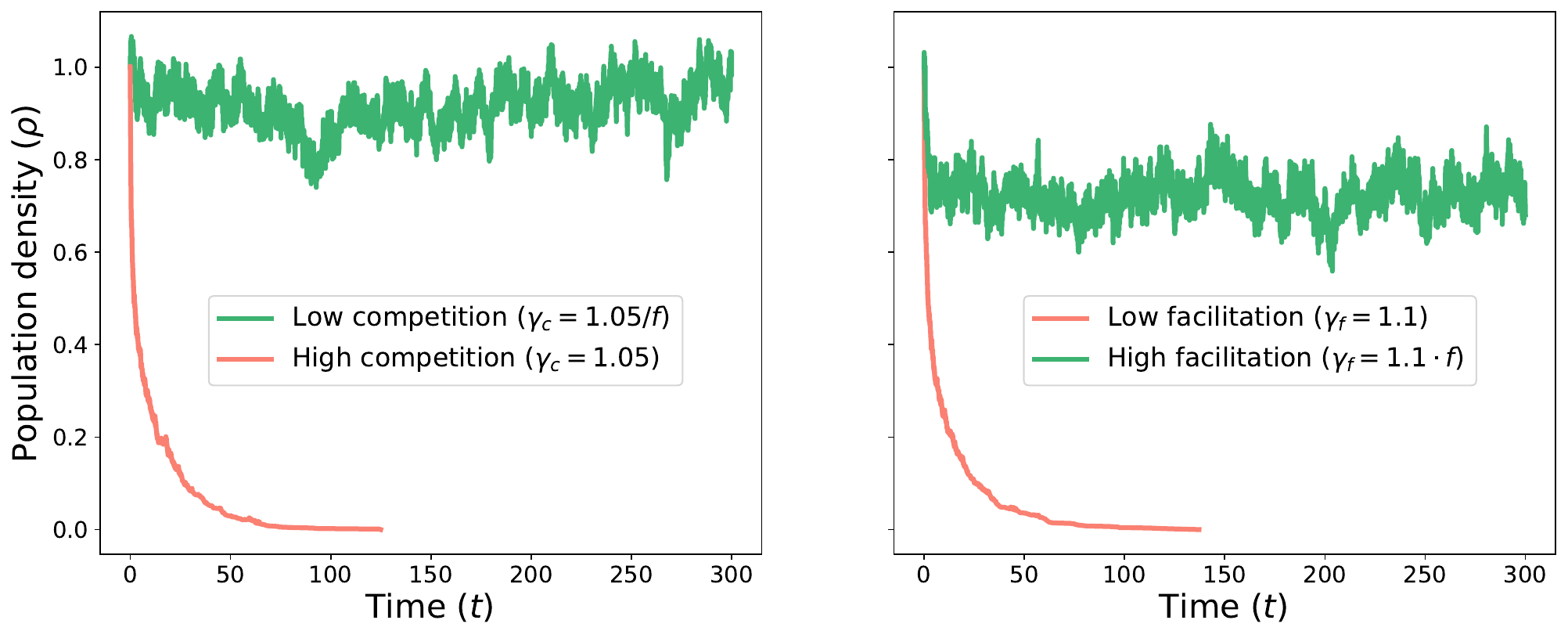}
        \caption{Example time series of a single realization of the Allee-based individual-based model in the absence of precipitation intermittency. Population density evolves under constant low (green) and high (red) competition or facilitation regimes. In the full stochastic model, precipitation events are implemented by switching between these two regimes through changes in the interaction kernel.}
        \label{fig:example_no_switching_allee}
    \end{figure}


    \begin{figure}[H]
        \centering
        \includegraphics[width=\linewidth]{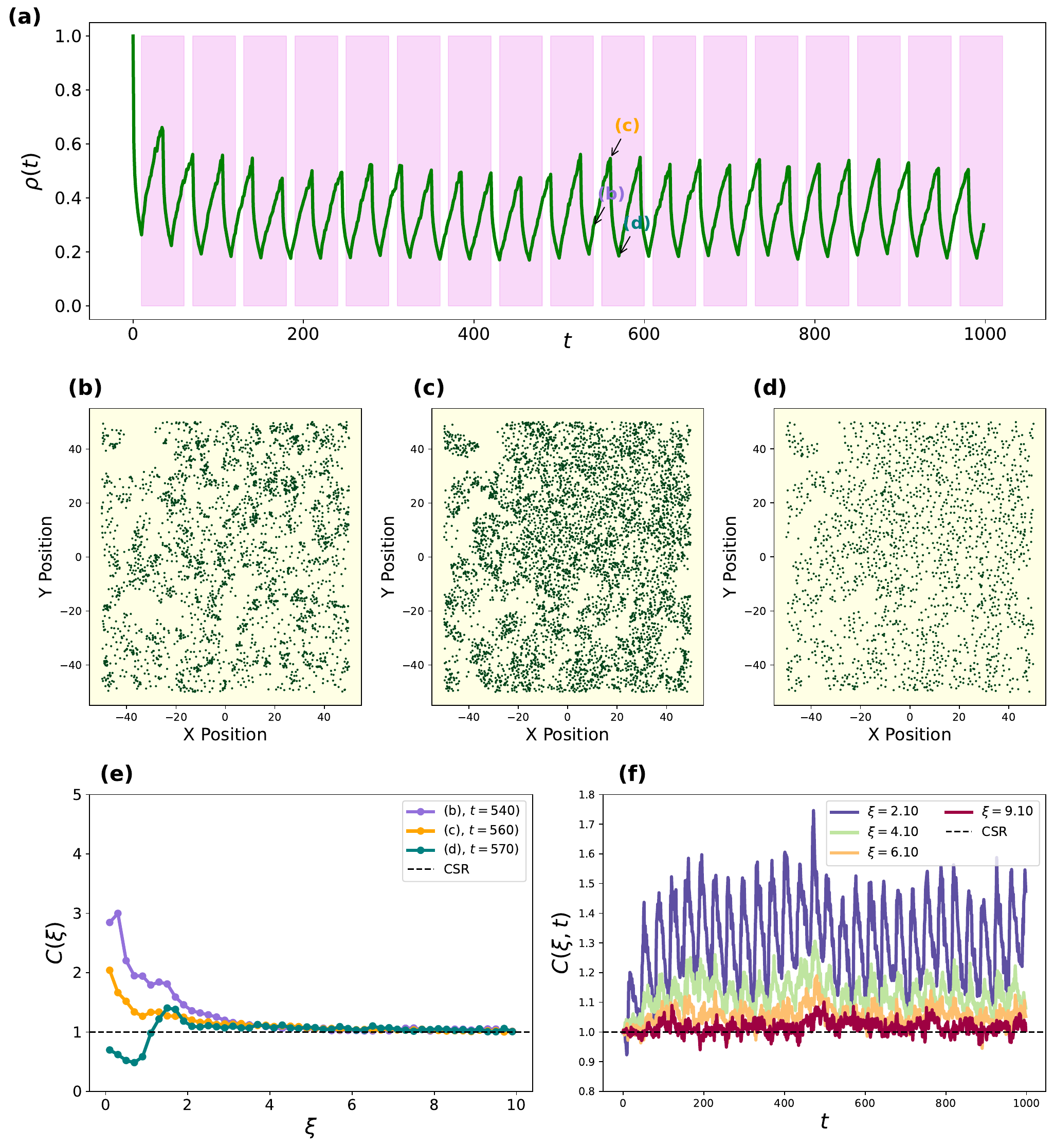}
        \caption{\textbf{Emergence and quantification of spatial clustering in a representative realization of the model with precipitation coupled to facilitation.} (a) Representative trajectory of population density as function of time with $d=25$ and $\lambda=0.1$. Pink regions correspond to time periods of enhanced facilitation due to precipitation events. (b)–(d) Snapshots of individual spatial configurations at successive times, showing the formation and reorganization of vegetation clusters. (e) Pair-correlation function computed at the times indicated in panels (b)–(d), compared with complete spatial randomness (CSR). (f) Time evolution of the pair-correlation function for different distances, illustrating the development and persistence of spatial structure relative to CSR.}
        \label{fig:big_system_facilitation}
    \end{figure}

    \begin{figure}[H]
        \centering
        \includegraphics[width=\linewidth]{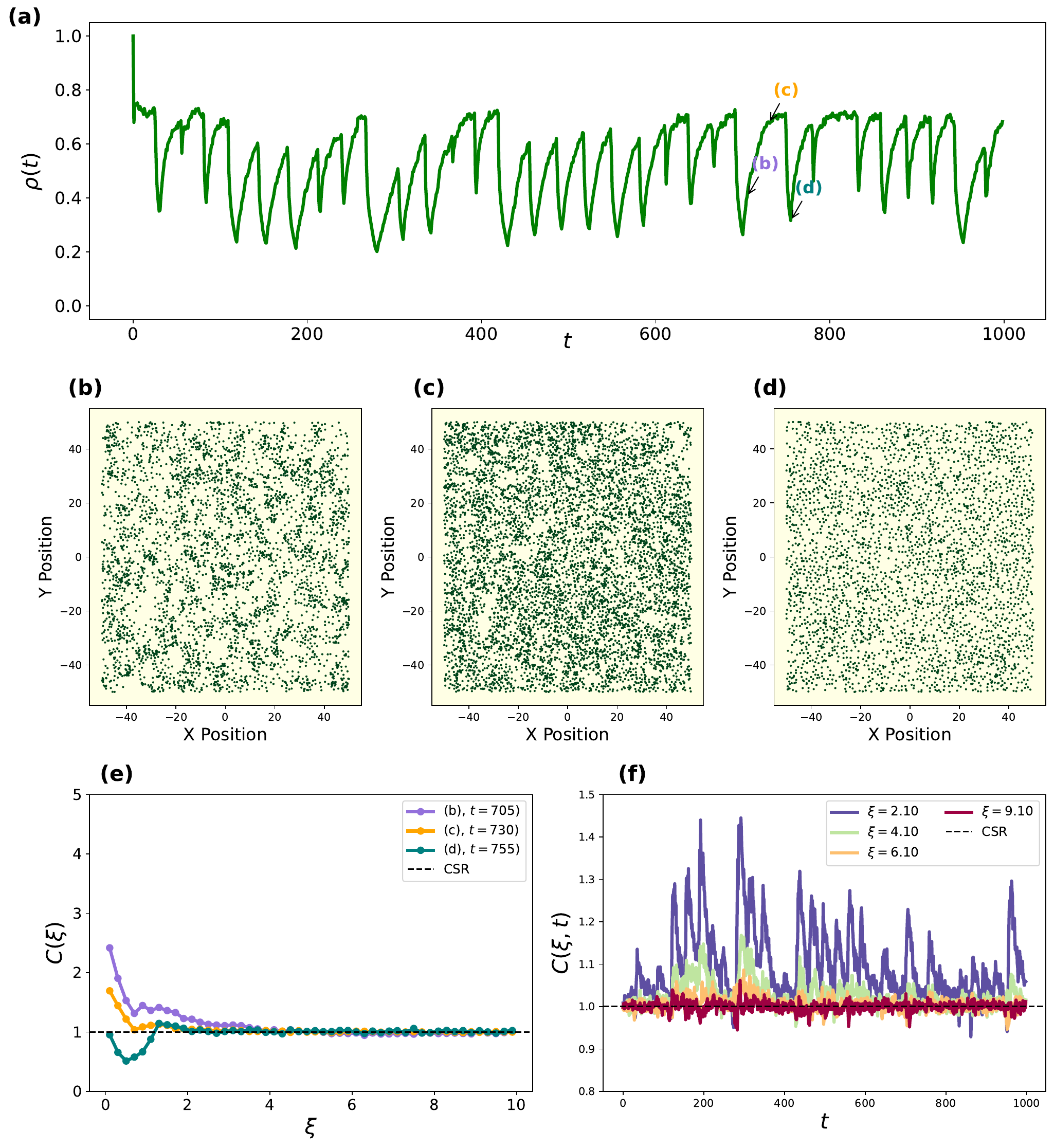}
        \caption{\textbf{Emergence and quantification of spatial clustering in a representative realization of the model with precipitation coupled to facilitation and stochastic precipitation events.} (a) Representative trajectory of population density as function of time with $d=25$ and $\lambda=0.2$. (b)–(d) Snapshots of individual spatial configurations at successive times, showing the formation and reorganization of vegetation clusters. (e) Pair-correlation function computed at the times indicated in panels (b)–(d), compared with complete spatial randomness (CSR). (f) Time evolution of the pair-correlation function for different distances, illustrating the development and persistence of spatial structure relative to CSR.}
        \label{fig:big_system_facilitation_stochastic_precipitation}
    \end{figure}

    \begin{figure}[H]
        \centering
        \includegraphics[width=\linewidth]{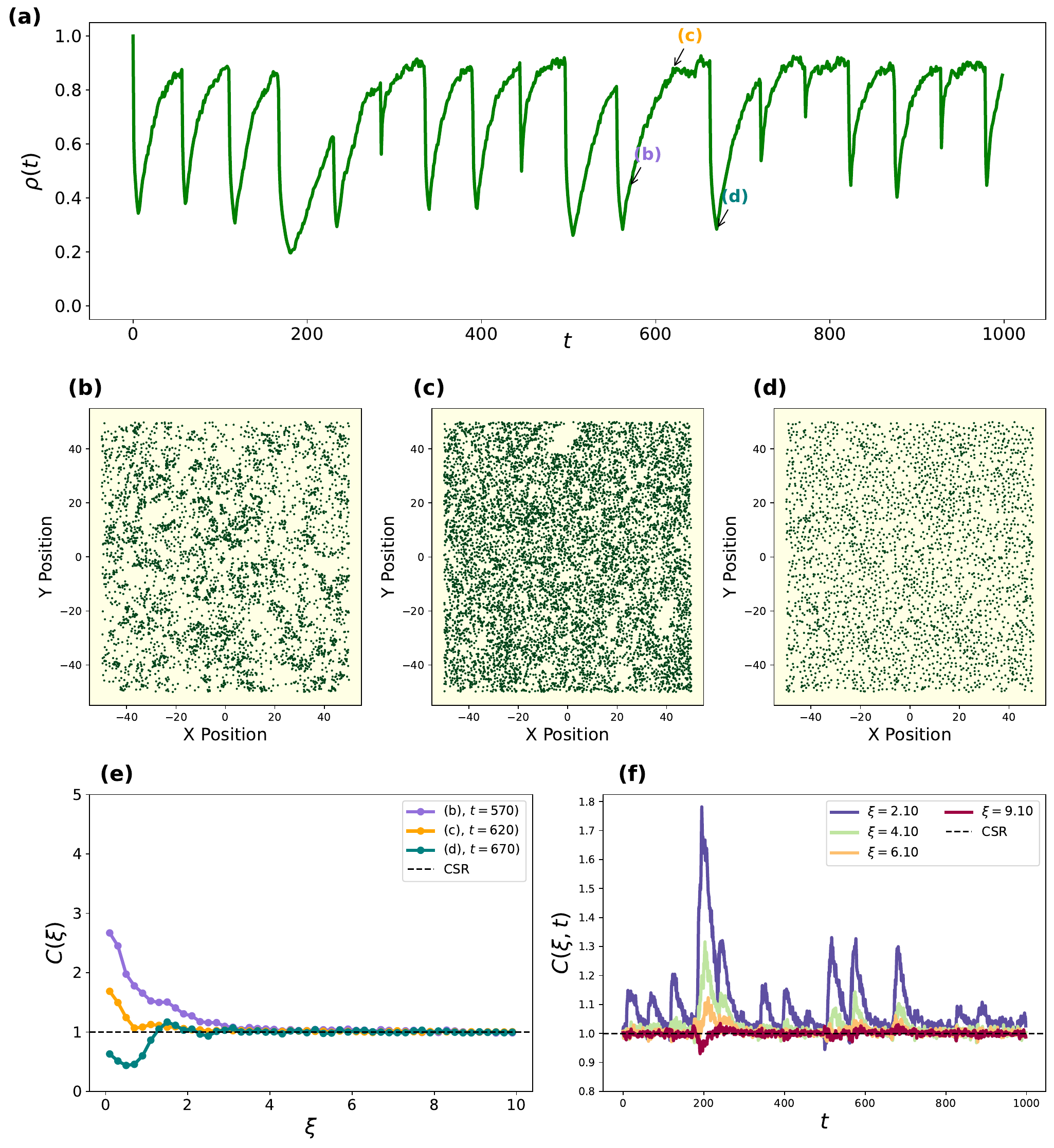}
        \caption{\textbf{Emergence and quantification of spatial clustering in a representative realization of the model with precipitation coupled to competition and stochastic precipitation events.} (a) Representative trajectory of population density as function of time with $d=50$ and $\lambda=0.2$. (b)–(d) Snapshots of individual spatial configurations at successive times, showing the formation and reorganization of vegetation clusters. (e) Pair-correlation function computed at the times indicated in panels (b)–(d), compared with complete spatial randomness (CSR). (f) Time evolution of the pair-correlation function for different distances, illustrating the development and persistence of spatial structure relative to CSR.}
        \label{fig:big_system_competition_stochastic_precipitation}
    \end{figure}

    \begin{figure}[H]
        \centering
        \includegraphics[width=0.8\linewidth]{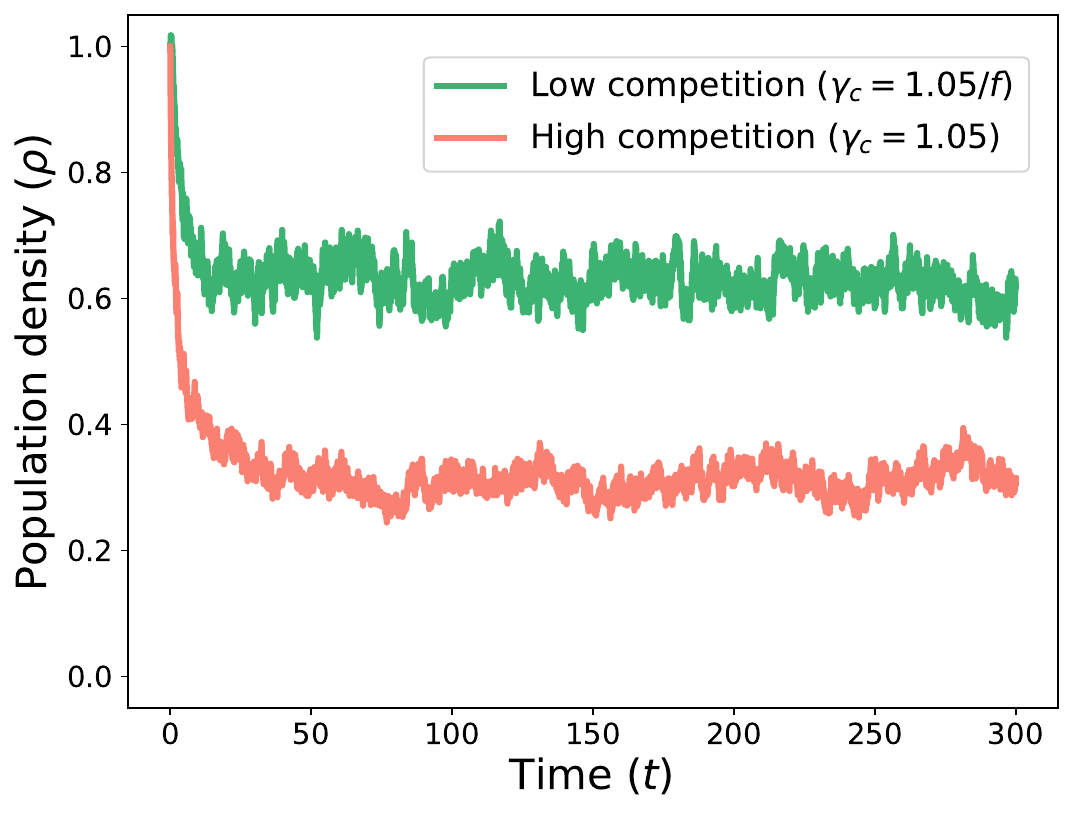}
        \caption{Example time series of a single realization of the logistic (non-Allee) model without precipitation intermittency, shown for constant low (green) and high (red) competition regimes. In simulations with precipitation events, the system alternates between these regimes by switching the competition kernel, analogously to the Allee-based model.}
        \label{fig:example_no_switching_logistic}
    \end{figure}

    \begin{figure}[H]
        \centering
        \includegraphics[width=0.8\linewidth]{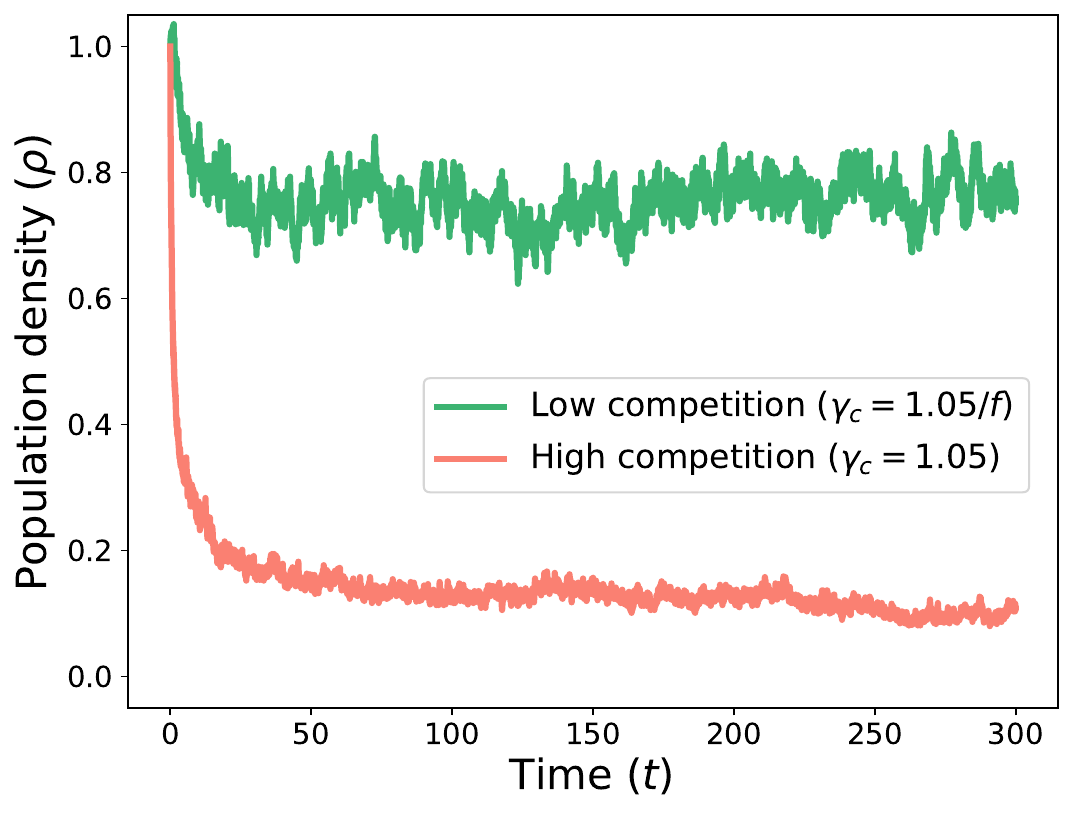}
        \caption{Example time series of a single realization of the logistic (non-Allee) model without precipitation intermittency, shown for constant low (green) and high (red) competition regimes with $\gamma_c=1.8$ and $\gamma_{c_-}=0.6$. In simulations with precipitation events, the system alternates between these regimes by switching the competition kernel, analogously to the Allee-based model.}
        \label{fig:example_no_switching_logistic_2}
    \end{figure}

    \begin{figure}[H]
        \centering
        \includegraphics[width=\linewidth]{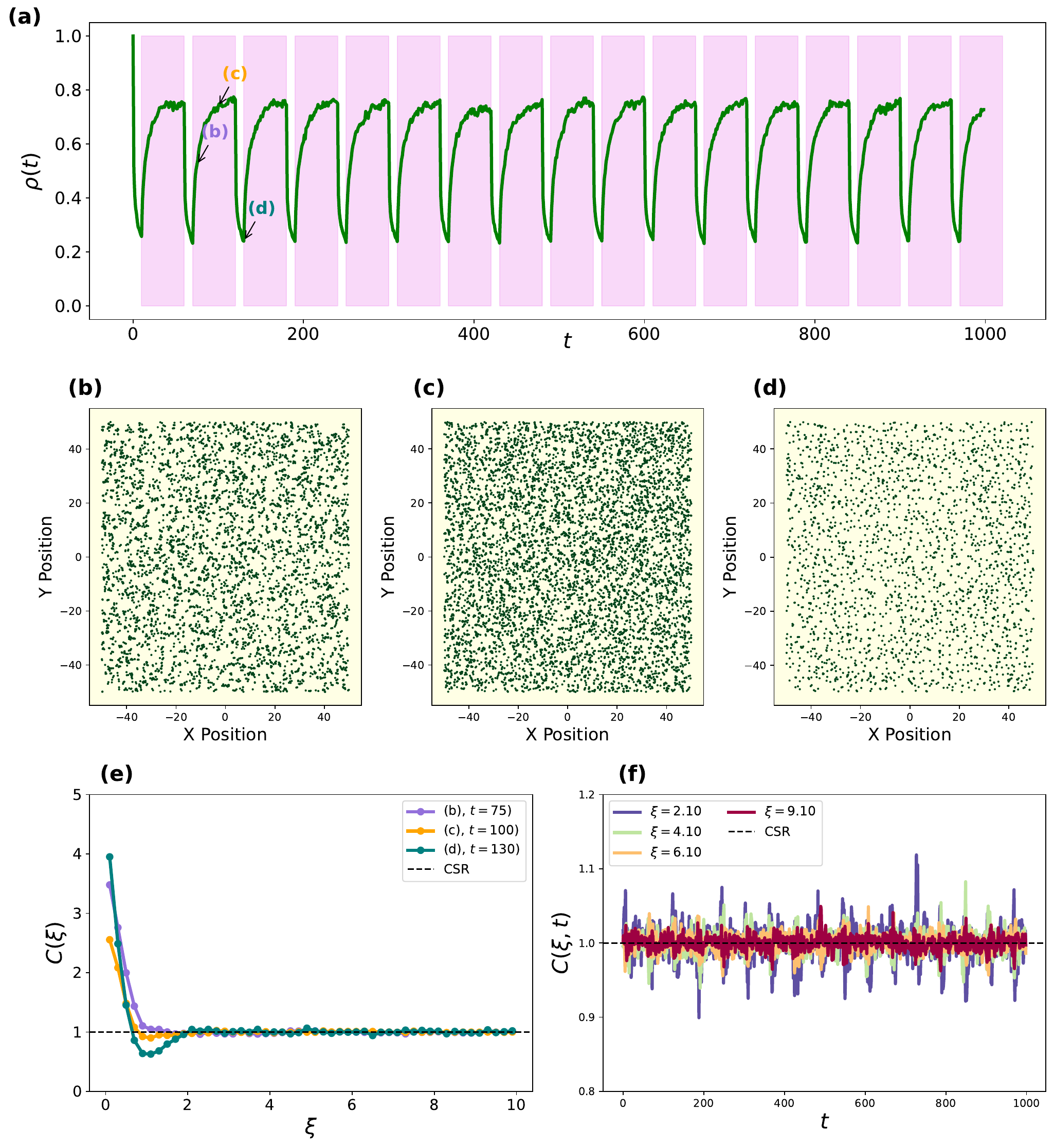}
        \caption{\textbf{Absence of large-scale clustering in the logistic model under intermittent precipitation with $\gamma_c=1.8$ and $\gamma_{c_-}=0.6$}. (a) Population density trajectory using a logistic model. (b–d) Spatial snapshots showing a near-uniform distribution of individuals; while small, transient groups appear due to short-range dispersal, no large-scale clusters emerge. (e) The PCF remains near 1 across most distances, suggesting little spatial correlation. (f) Time-resolved PCF confirms that without positive density-dependent feedback, vegetation cannot self-organize into structured patterns despite environmental fluctuations.}
        \label{fig:big_system_logistic_2}
    \end{figure}

    \begin{figure}[H]
        \centering
        \includegraphics[width=\linewidth]{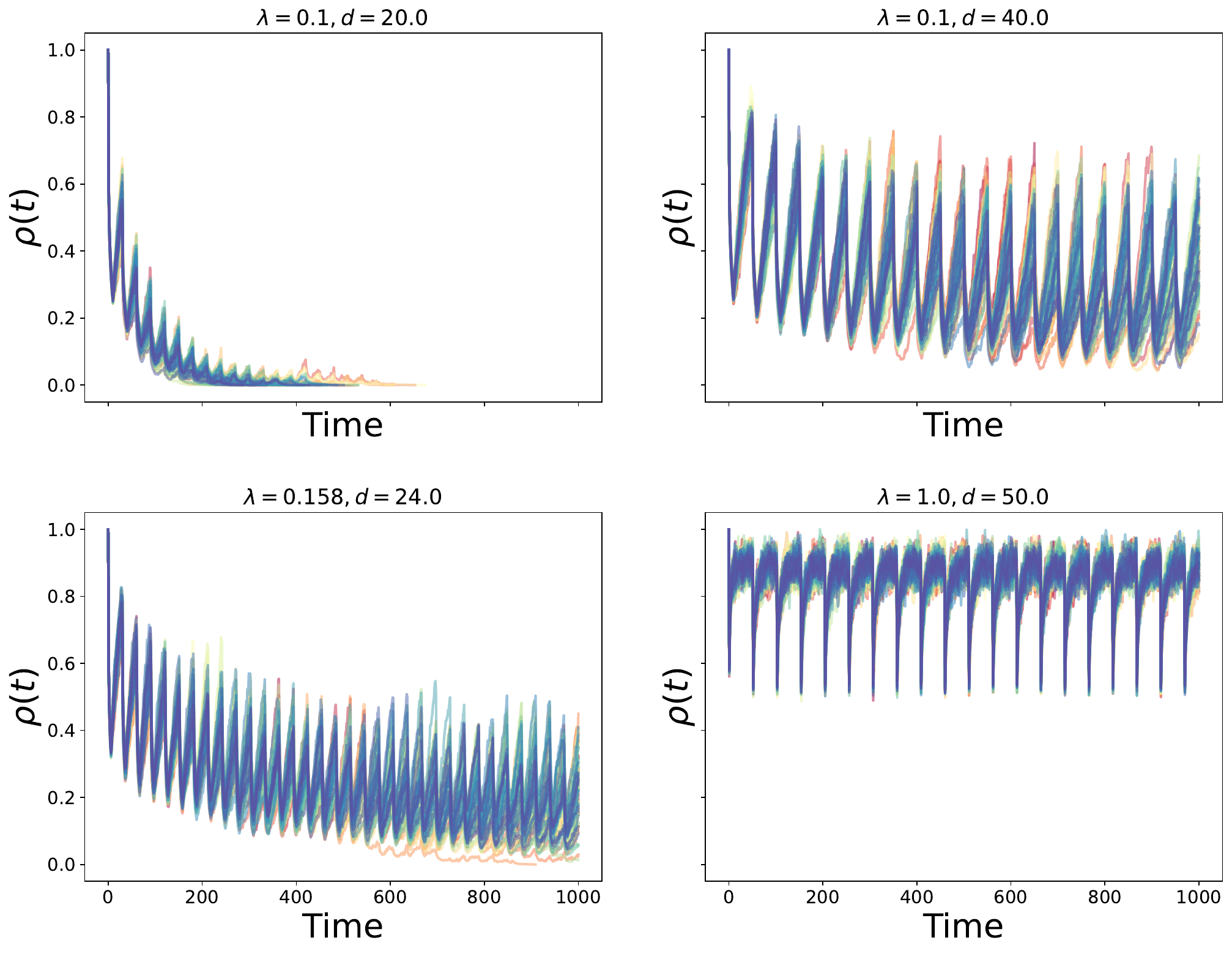}
        \caption{Individual realizations of the Allee-based model for different combinations of precipitation switching rate $\lambda$ and mean residence time $d$, with system size $L=50$. Precipitation intermittency is coupled to competition. The panels illustrate how changes in the frequency and duration of favorable periods affect population persistence and temporal variability.}
        \label{fig:individual_realizations_allee_competition}
    \end{figure}

    \begin{figure}[H]
        \centering
        \includegraphics[width=\linewidth]{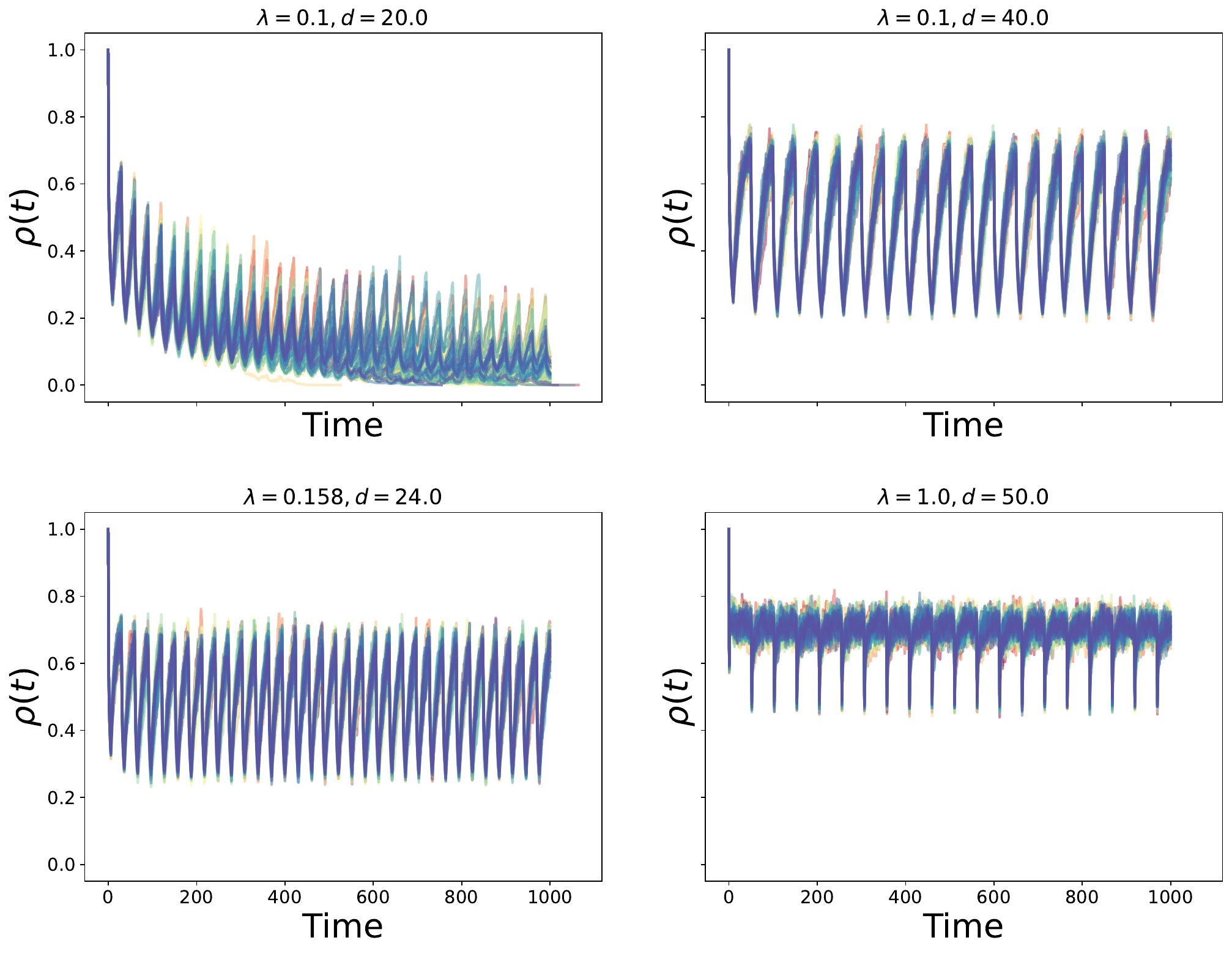}
        \caption{Individual realizations of the Allee-based model for the same parameter combinations as in \cref{fig:individual_realizations_allee_competition}, but with precipitation intermittency coupled to facilitation. Comparison with \cref{fig:individual_realizations_allee_competition} highlights that the qualitative effects of intermittency on persistence and variability are similar across coupling mechanisms, despite differences in the underlying interaction pathways}
        \label{fig:individual_realizations_allee_facilitation}
    \end{figure}

    \begin{figure}[H]
        \centering
        \includegraphics[width=\linewidth]{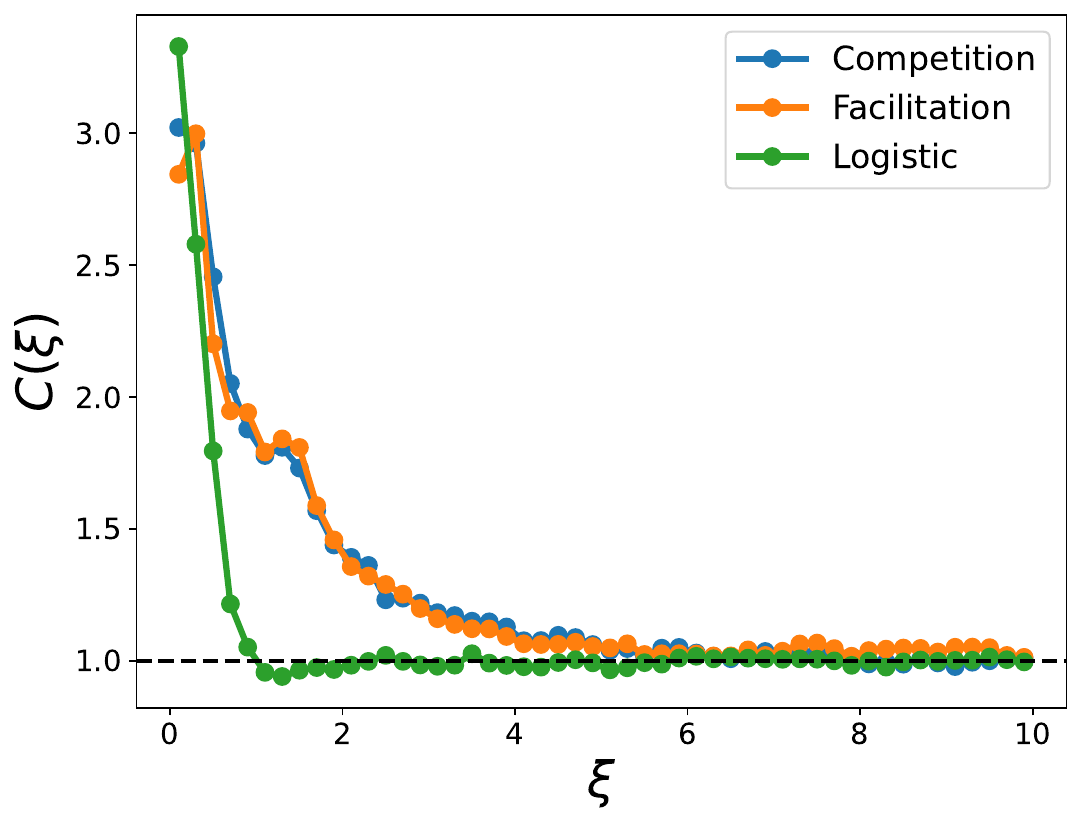}
        \caption{Pair-correlation functions (PCFs) obtained from the Allee model when precipitation modulates either competition (blue) or facilitation (orange), compared with the corresponding PCF of the logistic model (green). In the Allee model, the PCFs are nearly indistinguishable regardless of whether precipitation acts on facilitation or competition, indicating that spatial structure is robust to the specific coupling mechanism. By contrast, the logistic model exhibits a markedly different spatial signature, with a more rapid decay of correlations and a slight undershoot below unity, reflecting weaker clustering and the absence of positive density dependence.}
        \label{fig:comparison_PCF_allee_logistic}
    \end{figure}